%% file: paper.tex
\documentclass[letterpaper,twocolumn,10pt]{article}
\usepackage[twoside=true, head=13pt,
     paperwidth=8.5in, paperheight=11in,
     includeheadfoot=false, columnsep=2pc,
     top=1in, bottom=1in, inner=0.75in, outer=0.75in,
     marginparwidth=2pc,heightrounded]{geometry}

\usepackage{xcolor}
\usepackage{hyperref}
\hypersetup{
    colorlinks=true,
    linkcolor=red!50!black,  
    citecolor=red!50!black,  
    urlcolor=red!50!black    
}
\usepackage{graphicx}
\usepackage[tt=false, type1=true]{libertine}
\usepackage{mathptmx}
\usepackage{amssymb}
\usepackage[varqu]{zi4}
\usepackage{enumitem}
\usepackage{outlines}
\usepackage{graphicx} 
\usepackage{ftnxtra}
\usepackage{fnpos}
\usepackage[sort]{natbib}
\usepackage{stfloats}
\newsavebox{\bigimage}
\usepackage{pifont}
\usepackage{multirow}
\usepackage{morefloats}
\usepackage{amsmath}
\usepackage[normalem]{ulem}

\usepackage[subtle]{savetrees}
\usepackage[small,compact]{titlesec}
\usepackage[format=plain,labelfont=bf,font=small]{caption}
\usepackage{algorithm2e}
\usepackage{subcaption}
\usepackage{soul}
\setstcolor{red}
\usepackage{xurl}
\usepackage{appendix}

\definecolor{deepred}{rgb}{0.5,0,0}

\usepackage{lipsum}

\newcommand{\allnotes}[1]{}
\newcommand{\fixme}[1]{\allnotes{\bf\textcolor{red}{[#1]}}}
\newcommand{\radhika}[1]{\allnotes{\textcolor{magenta}{[Radhika: #1]}}}

\newcommand\paragraphb[1]{\noindent{\bf #1.}}

\usepackage{xspace}
\newcommand{\sysname}{\emph{REACT}\xspace}

\newcommand{\collsched}{collective pattern\xspace}
\newcommand{\collscheds}{collective patterns\xspace}
\newcommand{\irregular}{erratic\xspace}
\newcommand{\Irregular}{Erratic\xspace}
\newcommand{\cs}{CP}
\newcommand{\ML}{AI\xspace}

\title{Tuning Collective Patterns to Alleviate Congestion in Shared \ML Clusters}
\author{
    Eashan Gupta\\\emph{UIUC}\and
    Yongzhou Chen\\\emph{Meta}\and
    Apoorve Mohan\\\emph{IBM Research}\and
    Pavlos Maniotis\\\emph{IBM Research}\and
    Abdullah Kayi\\\emph{IBM Research}\and
    Radhika Mittal\\\emph{UIUC}}
\date{}

\begin{document}
\pagestyle{plain}
\maketitle
\begin{abstract}


Distributed \ML training involves recurring rounds of data exchange between multiple pairs of GPU nodes. Slowdown in even one flow due to congestion can cause the entire communication round to slowdown. Current approaches for evading congestion in \ML clusters assume global control over the entire workload (e.g. coordinating the schedule of all jobs) or assume infrastructural support (e.g. adaptive routing in switches). They are thus ill-suited in a shared cloud setting where \ML jobs belonging to one user can face external congestion from other users' jobs or background traffic beyond its own control. 
In this paper, we build a system, REACT, that tunes the recurring \emph{pattern} of data exchange between GPU nodes (known as communication collectives) in response to congestion. 
REACT works at the application (communication library) layer, where it detects congestion at runtime using readily available flow stats, and tunes the collective pattern to alleviate congestion -- changing the set of incident flows while retaining the semantics of information exchange (e.g. selecting which node aggregates data in an AllReduce tree). 
REACT requires no explicit support from the underlying network infrastructure and can be unilaterally deployed by individual users in a shared cloud setting. We prototype \emph{REACT} as a shim layer over NCCL, and evaluate it on a shared academic GPU cluster -- enabling \emph{REACT} improves communication performance (algorithm bandwidth) by $13\%-38\%$ under network congestion. Our simulations across a range of congestion scenarios further reveal up to $75\%$ performance improvement, highlighting the effectiveness of our approach. 

\end{abstract}
\sloppy

\input{intro-v4}
\input{background-v4}


\input{overview-v2}

\input{design-v2}

\input{implementation-v2}

\input{eval}

\input{discussion}

\input{conclusion}

\newpage
\bibliographystyle{plain}
\bibliography{hotnets24-template}
\clearpage
\appendix

\input{appendix}

\end{document}

%% file: intro-v4.tex
\section{Introduction}
\label{sec:intro}

Scaling machine learning (ML) and \ML training jobs to increasingly larger amounts of data and model sizes requires distributing them across multiple nodes. A typical distributed training job runs as a series of recurring \emph{epochs}, where each epoch is composed of a computation phase (where each node performs a local computation), followed by a communication phase (where relevant data, e.g., gradients and parameters, are exchanged among the nodes). The communication phase involves multiple flows (data exchange between multiple pairs of nodes). Due to synchronization barriers, slow down in even one flow (e.g. due to localized congestion) can slow down the entire communication phase. With the communication phase making up 10-90\% of the epoch duration~\cite{pipedream}, delays in communication directly translate to increase in epoch duration and the overall training time, and may even lead to underutilization of expensive GPU resources~\cite{meta-rdma, msft-gao}.
As a result, developing mechanisms to minimize (or evade) in-network congestion for \ML training jobs, in order to speed up their communication phase, has emerged as an active area of research~\cite{desensi2024swingshortcuttingringshigher, sccl, taccl, teccl, 
cassini, crux, adapcc, plink}. 

However, most prior work in this area focus on dedicated clusters that are used and managed by a single entity, and where one can assume 
complete knowledge and control over all jobs running in the cluster, and also control over the underlying network infrastructure.
An often overlooked setting is where distributed training jobs belonging to different users (or tenants) simultaneously run on a shared cluster (e.g. a large public cloud or a small-scale academic cluster). In such settings, network flows for a given training job can experience congestion from external sources (e.g. jobs belonging to other tenants, background data-transfers and storage tasks, etc)~\cite{msft-usenix}. Schemes designed for dedicated clusters (e.g. evading congestion by coordinating the temporal schedules of different jobs~\cite{cassini} or the paths and priorities assigned to their flows~\cite{crux}) are ill-suited in such shared settings: an individual tenant managing their own job has no control over other jobs belonging to a different tenant, or over the network infrastructure that is managed by the cloud operator. We also cannot rely on infrastructural mechanisms to dynamically route around congestion (e.g. adaptive routing in switches~\cite{adaptiverouting}), as they may or may not be supported by the shared cloud (the tenant has no visibility or control over it). 

So what can an individual tenant, running their own distributed \ML training job, do to evade congestion from external sources in a shared cloud? Motivated by this question, we build a system, \sysname (short for Runtime Epoch-Adaptive Collective Tuning) that tunes the recurring \emph{pattern} of communication among nodes in an \ML job in order to alleviate congestion -- changing the specific set of incident flows (source and destination nodes) while retaining the semantics of the required information exchange. REACT can be unilaterally deployed by an individual tenant to improve their own job's performance -- it works solely at the application (communication library) layer, and requires no explicit support from the network infrastructure and no control over external workload. 
It also makes no assumptions about the underlying network protocols (e.g. RDMA, TCP, etc), and is complementary to any transport-level congestion control mechanisms and routing strategies (e.g. ECMP, adaptive routing, etc) deployed by the cloud operator.

Before explaining how \sysname works, we provide some relevant context. Communication among the nodes in a distributed training job is specified through \emph{collective} operations that capture the high-level semantics of information exchange between nodes. Examples of such a collective operation includes ``AllGather'' where a piece of information (e.g. model weight or gradients) at each node must be sent to all other nodes, and ``AllReduce'' where the piece of information at all nodes must be aggregated and the result must be disseminated to all nodes. \ML training jobs commonly use a collective communication library or CCL (e.g. NCCL~\cite{nccl}, Gloo~\cite{gloo}, OpenMPI~\cite{mpi}) that executes the specified collective operation as a concrete sequence of data exchange steps among the participating nodes. We refer to this sequence of steps as the \emph{\collsched}. A given collective operation can be realized using different {\collsched}s. For example, an AllReduce between three nodes (N0, N1, and N2) can be realized using a tree pattern, where data from N0 and N1 is aggregated at N2 and the result is sent back to N0 and N1. The same operation can also be realized using a ring pattern where N0 sends its data to N1, which then aggregates it with its own data and sends the result to N2; N2 then aggregates the result with its data and sends the final result to N0 which then relays it to N1. 

The CCL computes the {\collsched}s once at the start of the job (typically based on parameters such as number of nodes, message sizes,  underlying transport protocols, etc) and uses the same pattern repeatedly in each epoch throughout the training run. 
Recent work proposes using additional information (e.g. profiled link characteristics, global knowledge of workload etc \cite{sccl,taccl,teccl,TacosWon_2024}) to optimally compute the collective pattern at the start of the job.  However, this initial collective pattern starts deviating from optimality as the network state changes over time (e.g. due to congestion caused by unforeseen arrival of external traffic). 

Our system, \sysname, tunes the \collsched at runtime in response to congestion. 
As a simple example, for the tree-based AllReduce with three nodes, if the incoming link at N2 is congested, rather than aggregating all data at N2, we can aggregate at N0 or N1. In the ring-based AllReduce example, if the path from N0 to N1 faces congestion, we can swap the positions of N1 and N2 in the ring, to avoid that path. In contrast to adaptive routing mechanisms that tune flow paths to evade congestion, \sysname changes the set of incident flows themselves (i.e. changes the source and destination nodes). It can therefore even help evade congestion on links that have no re-routing alternatives (e.g. the link from a ToR switch to a server \cite{homa}).

\sysname runs a feedback loop at the CCL level, where it collects the completion times of all flows after a small set of epochs, analyzes them to detect presence of congestion, and then tunes the \collsched in response to the detected congestion. 
Realizing such a feedback loop in practice requires tackling multiple challenges: 

(1) \emph{How do we reliably detect congestion using only flow-level stats?} For this, \sysname incorporates a mechanism for strategically comparing the completion time of various flows in a \collsched with one another, and across epochs, to identify flows that potentially experience congestion, without assuming any knowledge about the underlying network topology or link capacities. 

(2) Upon identifying the flows that experience congestion, \sysname iteratively picks such flows that lie on the critical path of the \collsched, and attempts to swap their source and destination nodes with other nodes in the \collsched in order to alleviate congestion. \emph{How do we determine which nodes can be swapped in a given \collsched, such that we retain semantic correctness?} For this, we outline a mechanism for computing viable \emph{transforms} for any given collective, that provides a set of viable nodes that any given node can be swapped with in the \collsched without violating semantic correctness. We further prune this set to eliminate nodes that would not help alleviate congestion.

(3) \emph{How do we ensure that REACT can update the \collsched fast, within a few epochs?} \sysname tests each transform to check whether it helps improve performance (and must be retained) or whether it should be reverted. Doing so one at a time by re-running the updated \collsched in the next epoch after each transform will increase convergence time. To speed up convergence, \sysname uses an analytical model to estimate the performance impact of a transform without actually running the updated \collsched. This allows \sysname to iterate through a batch of multiple transforms within the same epoch.
We provide more details on each of these design aspects in \S\ref{sec:design}. 



Note that \sysname can only respond to congestion that lasts for a few training epochs (i.e. for a duration of a few hundreds of milliseconds), e.g. from long-running data transfer tasks or other competing \ML jobs. It is not meant to react to ephemeral congestion that dissipates within an epoch. Further note that our contributions are orthogonal and complementary to how the initial \collsched is generated in the first place. 
Finally, it may not always be possible to eliminate \emph{all} congestion by swapping nodes -- \sysname is a best-effort system that attempts to alleviate congestion to the best extent possible. 

We prototype \sysname, implementing it as a shim layer over PyTorch and NCCL.  We evaluate it on a national academic shared GPU cluster, and show how enabling \sysname 
improves the communication time by 13\%-38\% under different degrees of congestion. 
We further evaluate \sysname by simulating a wide variety of congestion and network failure scenarios in ns3~\cite{ns3} to show up to $75\%$ improvement in algorithm bandwidth. Our evaluation spans a variety of popularly used {\collsched}s: NCCL's dual binary tree based AllReduce~\cite{nccltreeallreduce, tree-allreduce}, Ring AllReduce~\cite{ring-allreduce, all-coll-algos}, and AllGather with recursive doubling~\cite{nccl-collectives-2,all-coll-algos}.

%% file: background-v4.tex
\section{Background}
\label{sec:background}


\paragraphb{Repeated rounds of information exchange} A typical \ML training job runs over multiple rounds or epochs, where model weights and parameters are repeatedly updated using iterative gradient descent. To scale training, the job is typically split between several nodes (or GPUs) by sharding the training data, model, and/or the tensors~\cite{megatron, deepspeed,NIPS2014_186b3d04,shoeybi2020megatronlmtrainingmultibillionparameter}. 
Each epoch in such a distributed training job therefore comprises of a combination of local computation at each node, and communication phases where relevant data (gradients and updated weights) are exchanged among nodes. With each epoch repeating the same gradient descent logic, the same high-level information exchange between nodes is repeated in each epoch. These epochs therefore provide a useful time granularity for a reactive feedback loop, which is unique to \ML/ML training, and is exploited by our system.


\paragraphb{Information exchange specified via collective operations} The repeated information exchange between the participating nodes is specified at a high-level through collective operations (e.g. AllGather, AllReduce, etc, as mentioned in \S\ref{sec:intro}).  
Application-level libraries such as PyTorch~\cite{pytorch} or Tensorflow~\cite{tensorflow} use the model specification and the set of participating nodes to generate a directed acyclic graph (DAG) of computation and communication tasks, schedule them across devices, and specify the communication requirements using collective operations. For the latter (i.e. for carrying out the communication tasks), these application libraries invoke a collective communication library (CCL), such as NCCL~\cite{nccl}, Gloo~\cite{gloo} and OpenMPI~\cite{mpi}, through standard collective APIs. 

\paragraphb{Collective operations realized via {\collsched}s} 
Given the collective operation, CCL computes the specific collective algorithm to realize the operation, i.e. how the data should be chunked up, and the concrete series of message exchanges among participating nodes. We refer to this series of message exchanges as the \emph{\collsched}. A given collective operation can be realized through several different patterns (all resulting in the same high-level information exchange needed at each training epoch). For example, AllReduce can be accomplished using a tree pattern or a ring pattern (as exemplified in \S\ref{sec:intro}), as well as other patterns. AllGather can be accomplished through broadcast by each node, or through a more structured recursive doubling pattern~\cite{all-coll-algos}. We provide more details about these popularly used {\collsched}s in Appendix~\ref{sec:nccl-collectives}, and briefly discuss how CCLs select {\collsched}s in the next section. 

The CCL executes the collective operation by invoking the underlying transport (PCIe or NVLink for intra-host communication, RDMA or TCP/IP for inter-host communication, etc) and setting up the message exchanges between the corresponding source-destination pairs, as determined by the \collsched. The \collscheds typically have a hierarchical structure, where the intra-host data exchange is handled first, followed by inter-host data exchange. In this work, we focus on the inter-host component of collective communication (discussing the intra-host aspects in \S\ref{sec:discussion}).

\section{Related Work}

\subsection{Computing Collective Patterns}

\noindent\textbf{Heuristic-based Collective Patterns.}
Using communication collectives to exchange data is an extensively studied field in HPC and ML communication, and different heuristics have been developed to compute collective patterns for specific network topologies and stacks~\cite{desensi2024swingshortcuttingringshigher,nccltreeallreduce, tree-allreduce,ring-allreduce, all-coll-algos,nccl-collectives-2}\fixme{cite others}. For instance, NCCL, the most widely used CCL for distributed training, 
maintains a fixed set of patterns for each collective operation (e.g. a dual binary tree and a ring for AllReduce), and selects the lowest cost pattern for the given setting, where cost is computed based on the message size, number of nodes/devices (GPUs, NICs) and pairwise link bandwidth and latency (that are hardcoded for each underlying protocol -- RDMA, TCP, GPUDirect, SHARP, NVLink, PCIe, etc). 

\noindent\textbf{Solver-generated Collective Patterns.}
Several recent works (e.g.~\cite{sccl,taccl,teccl,TacosWon_2024}) propose finding the optimal collective pattern by modeling the underlying network and workload characteristics in solvers like Z3 and Gurobi, with the objective of reducing the overall communication time. While such approaches promise more optimal outcomes than heuristics, they are computationally expensive to run repeatedly. Additionally, they require complete knowledge of link characteristics and all workloads in the cluster to compute the optimal \collsched, which is not always possible.

Both of the above approaches generate a collective pattern once at the start of the job, and use it repeatedly over the subsequent epochs. Our work takes such a pre-computed \collsched as input, and minimally tweaks it at runtime in order to react to congestion. 

\noindent\textbf{Adapting Collective Patterns.} There have been a few proposals to adapt communication collectives at runtime. Plink~\cite{plink} periodically probes all pairwise network paths, and uses the resulting bandwidth and delay information to choose the root and leaf nodes for AllReduce operations realized via two-level hierarchical trees. 
However, such constant probing is costly and cannot be done at small enough timescales (especially when frequent re-profiling is required to capture the variance that may arise from probabilistic ECMP hash collisions).
Moreover, Plink adaptation is restricted to how the root and leaf nodes are chosen in the hierarchical tree. In contrast, our goal is to provide a more general reaction to congestion for any given \collsched. 

Another work, AutoCCL~\cite{autoccl} also tunes collective communication at runtime, but focuses on only tuning pre-defined NCCL parameters  (e.g. the chunk size, number of channels, threads, etc.) without changing the intrinsic communication pattern itself. 

Another closely related work is AdapCC~\cite{adapcc}, that also periodically probes all pairwise network paths at coarse timescales of every five hundred epochs, and uses a solver to recompute the \collsched based on the updated information about network delays (running such a solver at smaller timescales is prohibitively expensive).  
In contrast, our system, \sysname, works at a much smaller timescale of a few epochs that is complementary to AdapCC -- \sysname can use the infrequently generated collective patterns by AdapCC as inputs, and minimally tweak them to provide an order of magnitude faster reaction to congestion.  
\radhika{as we demonstrate in \S\ref{sec:eval}.}

\subsection{Minimizing Network Delays in Distributed Training}
\label{subsec:react-to-cong}


The standard way to react to network congestion is for the congestion control algorithm used by the underlying transport (e.g. TCP \cite{tcpcongestioncontrol}, RDMA \cite{dcqcn}, or customized transport~\cite{meta-rdma,swift,falcon,cray}) to kick in and reduce the sending rate of the flow. However, this still increases the flow completion time, and thereby the training time. Therefore, several solutions have been proposed to evade congestion during \ML training.

\noindent\textbf{Routing-based Solutions.}
One of the proposed solutions is to avoid congested paths by rerouting flows using infrastructural support, e.g. adaptive routing in switches~\cite{adaptiverouting}, or source routing via port changes in the hardware transport or the NIC (that  determine ECMP hashes)~\cite{crux,plbSigcomm2022,bonato2026spritzpathawareloadbalancing,mcclure2026loadbalancingaitraining,srv6-rfc8986}. Another alternative is to avoid creating hotspots in network links by leveraging switch support to uniformly spray packets~\cite{dcp-ho-trimming,reps,khashab2026highspeednetworkinggigascaleai,mrc}. Such infrastructure-level solutions are beyond the control of individual tenants using a public cloud cluster (they may or may not be supported or enabled). Moreover, re-routing cannot help when congestion happens on an unavoidable link with no alternatives (e.g. the link from a ToR switch to a server \cite{homa}). 

\noindent\textbf{Coordinated Scheduling.} Another set of solutions minimize the effects of congestion among competing \ML jobs by coordinating their spatial or temporal schedules. For instance, Cassini~\cite{cassini} develops a global scheduler that computes the temporal schedule of all jobs in the cluster to minimize overlap in their communication phases. Crux~\cite{crux} coordinates the path selection and priorities across all jobs to reduce the communication time that would result in idle GPU. Both of these approaches assume control over the entire workload (which is beyond the control of an individual tenant in a shared cloud), with Crux additionally assuming infrastructure support for in-network telemetry, source routing, and prioritization. 

Another body of work attempts to reduce communication delays using a best-effort co-location strategy to place a given job on closeby machines~\cite{themis,pollux} -- these work are orthogonal and complementary to \sysname, which attempts to alleviate inter-host congestion for a given workload placement. 

\noindent\textbf{Application-Integrated Approaches.} 
Another orthogonal and complementary body of work attempts to minimize the effects of communication delays by increasing the overlap in computation and communication phases~\cite{syndicate,ByteScheduler,poseidon}, 
or reducing the data to be communicated via compression or quantization ~\cite{syndicate,ByteScheduler,poseidon}, or dropping low-information packets during communication to reduce tail latencies~\cite{optireduce,mlt}. These work require tighter integration with the \ML application and are not semantically transparent. \sysname, in contrast, works at the communication collective library layer, without requiring any modifications in the overlying application (or the underlying transport and network infrastructure). 





%% file: overview-v2.tex
\section{Overview}
\label{sec:overview}

\vspace{-3pt}
\subsection{Key Design Considerations}
\vspace{-2pt}




\paragraphb{Target Scenarios} 
We design \sysname for shared GPU clusters, where distributed training jobs from one user (or tenant) can face external congestion from competing distributed \ML workloads from other tenants, large file transfers, or even the other communication tasks within the same training job like check-pointing or copying training data from storage servers to GPUs. These congestion scenarios can last for several hundreds of milliseconds or even longer~\cite{congPatternsMetaIMC2025}, providing time to react over timescales of a few training epochs.

Our targeted settings may range from large high-end clusters in the cloud to smaller-scale shared academic clusters. These clusters may vary in network topology (large-scale Clos networks connecting thousands of servers vs small-scale star topology on a single rack), the underlying network protocols (different variants of RDMA, TCP/IP, or proprietary solutions~\cite{adaptiveroutingwhitepaper,falcon,swift,cray}, ECMP vs adaptive routing, etc), server configurations, and so on. These infrastructural aspects are determined by the cluster operators, with individual tenants having limited or no control over them. We design \sysname to work across all of these various infrastructural settings and underlying network protocols. \sysname can be unilaterally deployed by individual tenants or users to alleviate congestion for their own \ML training job within a few training epochs, irrespective of how the underlying network is managed by the cluster operators. This requires us to tackle the following design challenges:



\noindent
\textbf{Challenge 1: Only application-layer changes.} 
We cannot assume any network support, or make low-level infrastructural changes (such as tuning the network paths traversed by flows~\cite{crux,adaptiverouting,bonato2026spritzpathawareloadbalancing}), as these are beyond the control of individual tenants. To enable individual tenants to unilaterally deploy \sysname, it works solely at communication collective library (CCL) level. Specifically, \sysname uses flow-level stats readily available at the CCL to detect congestion. It then reacts to congestion by changing the \collsched that is executed by the CCL. 
Deployment options for \sysname can range from incorporating \sysname logic into existing CCLs (NCCL, Gloo, OpenMPI, etc) or deploying \sysname as a separate higher-level library over an existing CCL (as done in our prototype implementation in \S\ref{sec:impl}). 

\noindent
\textbf{Challenge 2: No global cluster information.} We assume that each training job runs independently and has no visibility into other concurrently running jobs (potentially from different tenants) or the global network state. This restricts \sysname to only using information that is locally available at the CCL for a given job. \sysname runs a reactive feedback loop system based on this local information. 
This is in contrast to solutions such as Cassini~\cite{cassini} that assumes knowledge about all jobs running in the cluster to co-optimize their schedule. 



\noindent
\textbf{Challenge 3: Low overhead.} We design \sysname to be a low cost solution that users can easily deploy, and that can react within timescales of a few epochs. To that end, we eschew the use of explicit link profiling for determining network conditions (as in~\cite{plink, adapcc}), instead relying on flow-level stats readily available at the CCL. Moreover, we choose to use simple heuristics to minimally tweak the given \collsched in response to congestion rather than using expensive techniques to recompute the entire pattern (as in AdapCC~\cite{adapcc}). Tweaking the \collsched minimally also helps minimize the overheads of updating the pattern (which can require setting up new connections) and further try to maintain their desirable properties (like latency or bandwidth optimality).

\subsection{\sysname in a nutshell}


\sysname uses a feedback loop to explore and tune the given \collsched in response to congestion, based on the current network state. At the end of each training epoch, \sysname collects the flow completion times (FCTs) of individual flows from each node that participates in the \collsched, and sends them to the leader (a randomly selected node among all participating nodes). After a small batch of epochs (set to 6 in our implementation), the \sysname logic running at the leader analyzes the FCTs to detect presence of congestion and to identify potential sources of congestion. It then modifies the \collsched by applying a series of \emph{transforms} in order to circumvent congestion and improve performance. The modified \collsched is then disseminated to all participating nodes and applied for the next batch of epochs. This process repeats until no further congestion is detected, or no further changes help improve performance.
Our transform logic is designed to ensure that all modifications made to the \collsched are \emph{safe} i.e. any modification preserves the communication semantics expected by the application, resulting in the same logical data exchange as required by the collective. 
\S\ref{sec:design} provides details of how the feedback loop works, how \sysname detects congestion, and how it transforms the \collsched. 

\subsection{Illustrative Example}
\label{subsec:example}

\begin{figure*}[t!]
    \centering
    \sbox{\bigimage}{%
        \begin{subfigure}[t]{.395\textwidth}
            \centering
            \includegraphics[width=\textwidth]{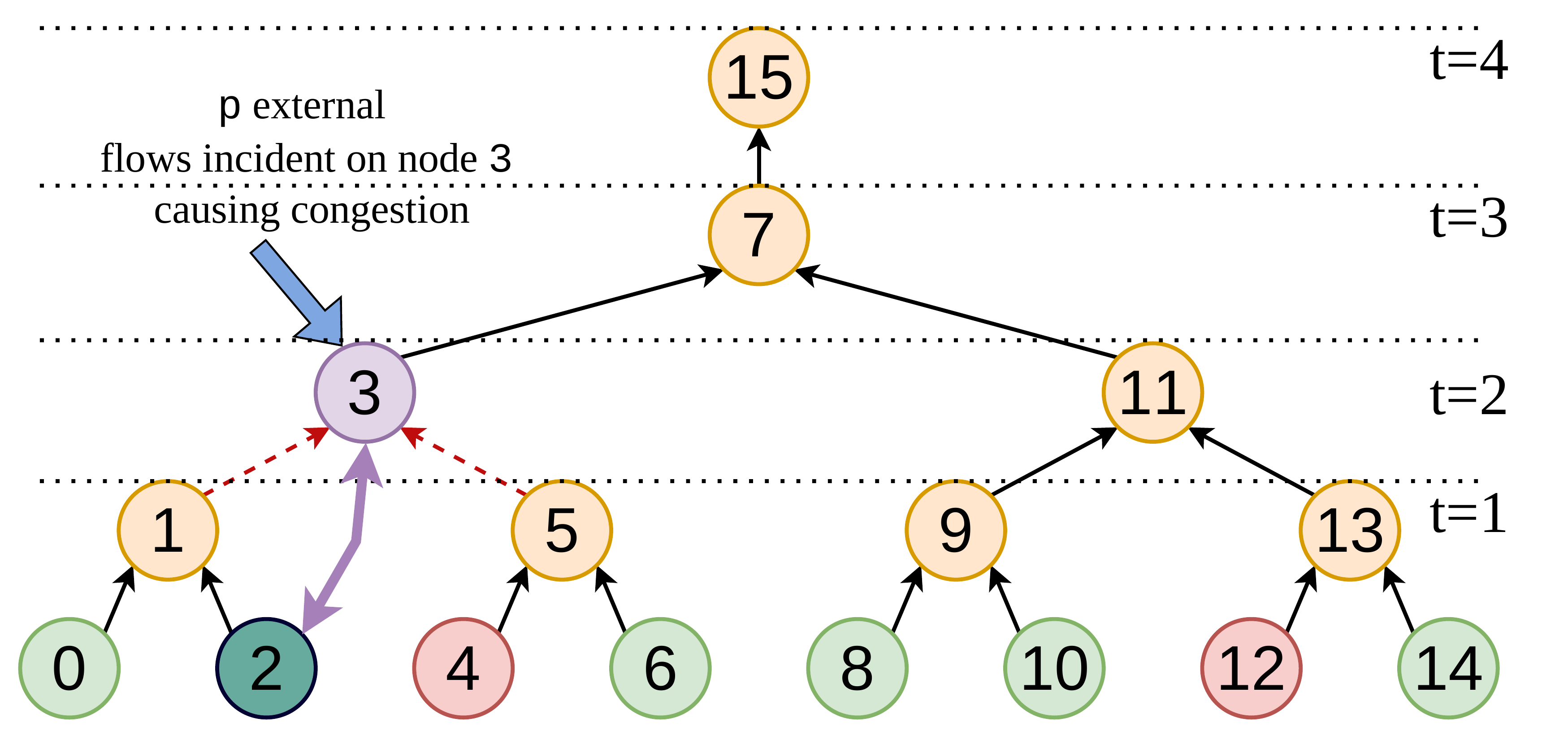}
            \caption{Even leaf-node tree (all leaf nodes are even ranks)}
            \label{subfig:eventree}
        
            \vspace{0pt}
        \end{subfigure}%
        ~
        \begin{subfigure}[t]{.395\textwidth}
            \centering
            \includegraphics[width=\textwidth]{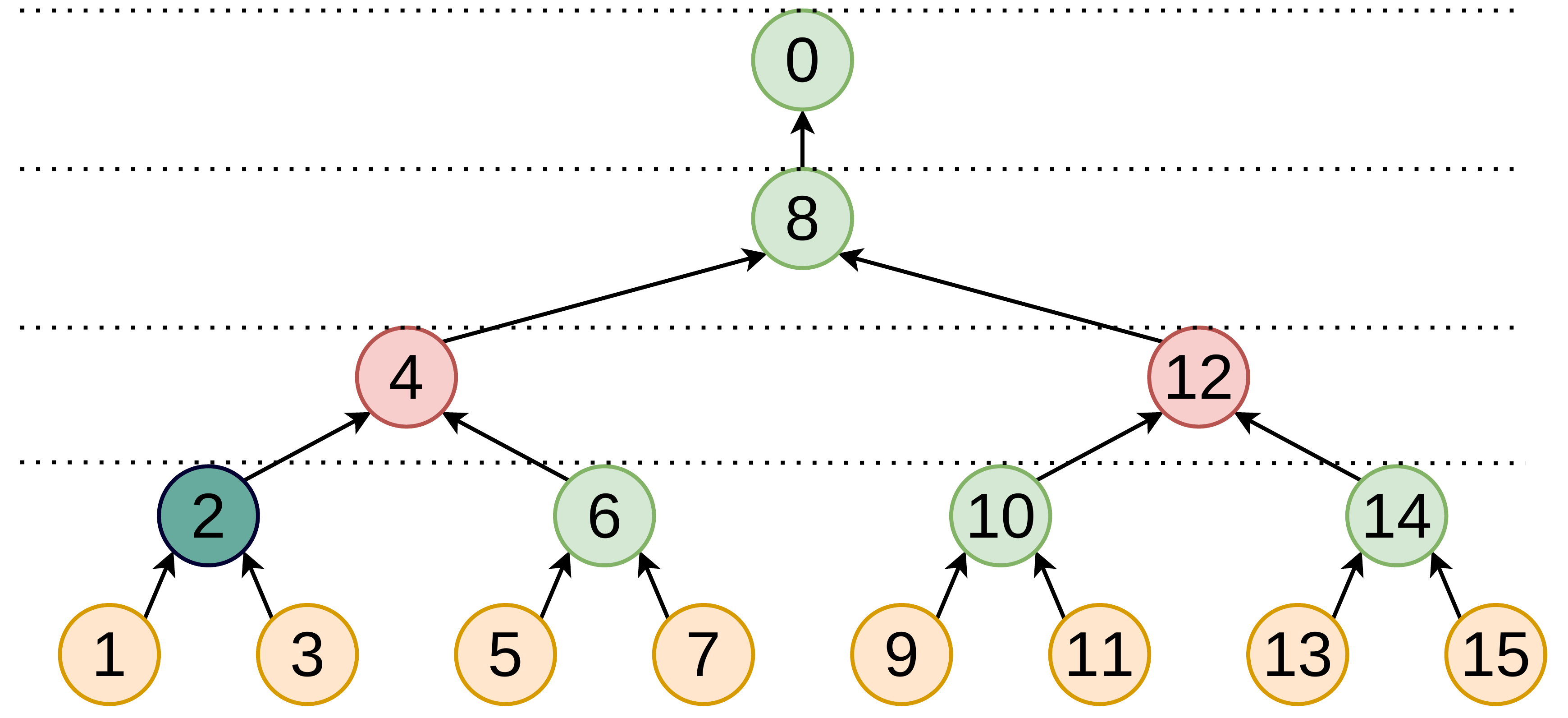}
            \caption{Odd leaf-node tree (all leaf nodes are odd ranks)}
            \label{subfig:oddtree}
            
            \vspace{0pt}
        \end{subfigure}%
    }
    
    \usebox{\bigimage}
    \begin{minipage}[b][\ht\bigimage][s]{.16\textwidth}
        \begin{subfigure}{\textwidth}
            \centering
            \includegraphics[height=1.32cm]{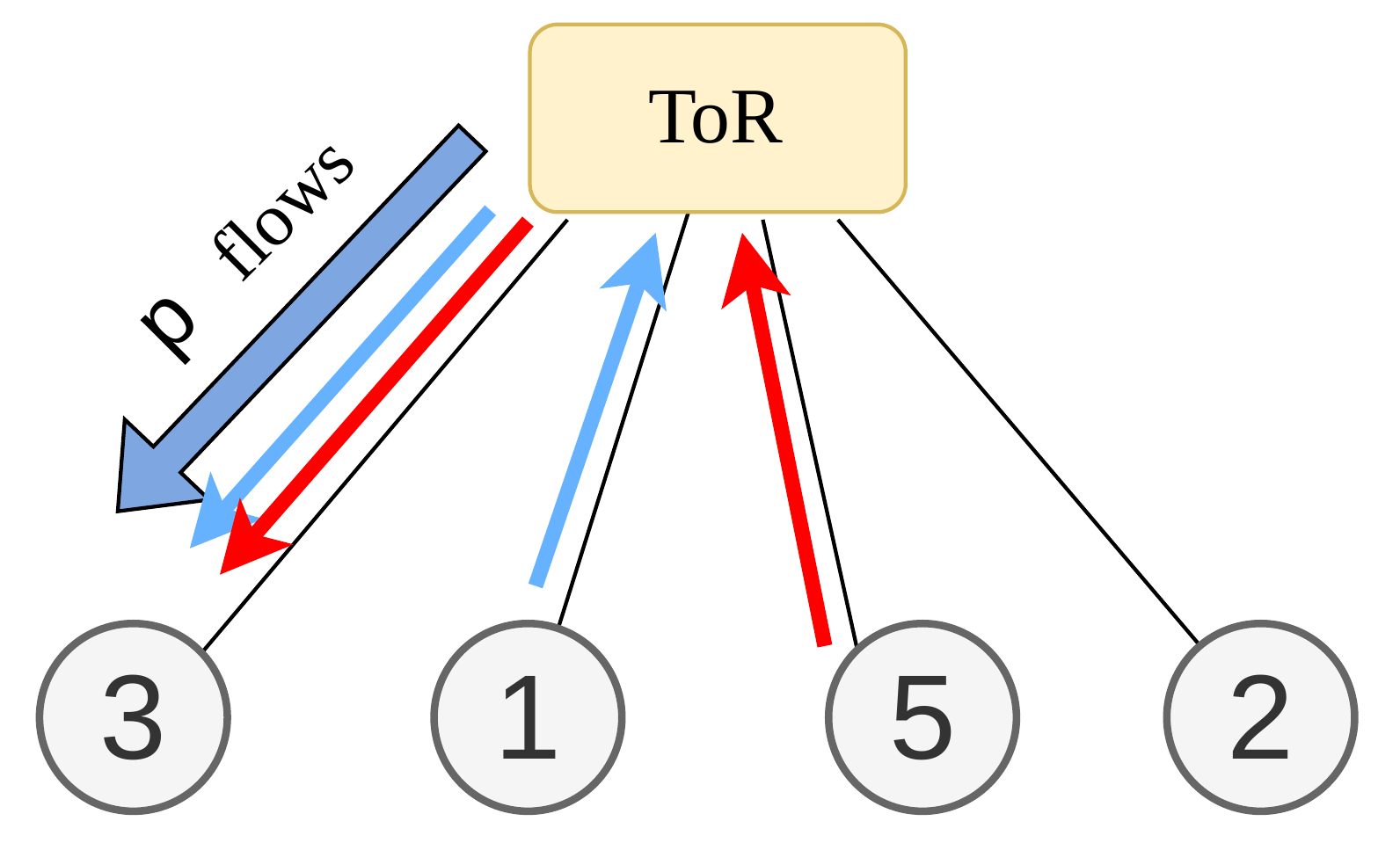}
            \caption{Before replacement
            }
            \label{subfig:beforereplacement}
        \end{subfigure}%
        \vfill
        \begin{subfigure}{\textwidth}
            \centering
            \includegraphics[height=1.32cm]{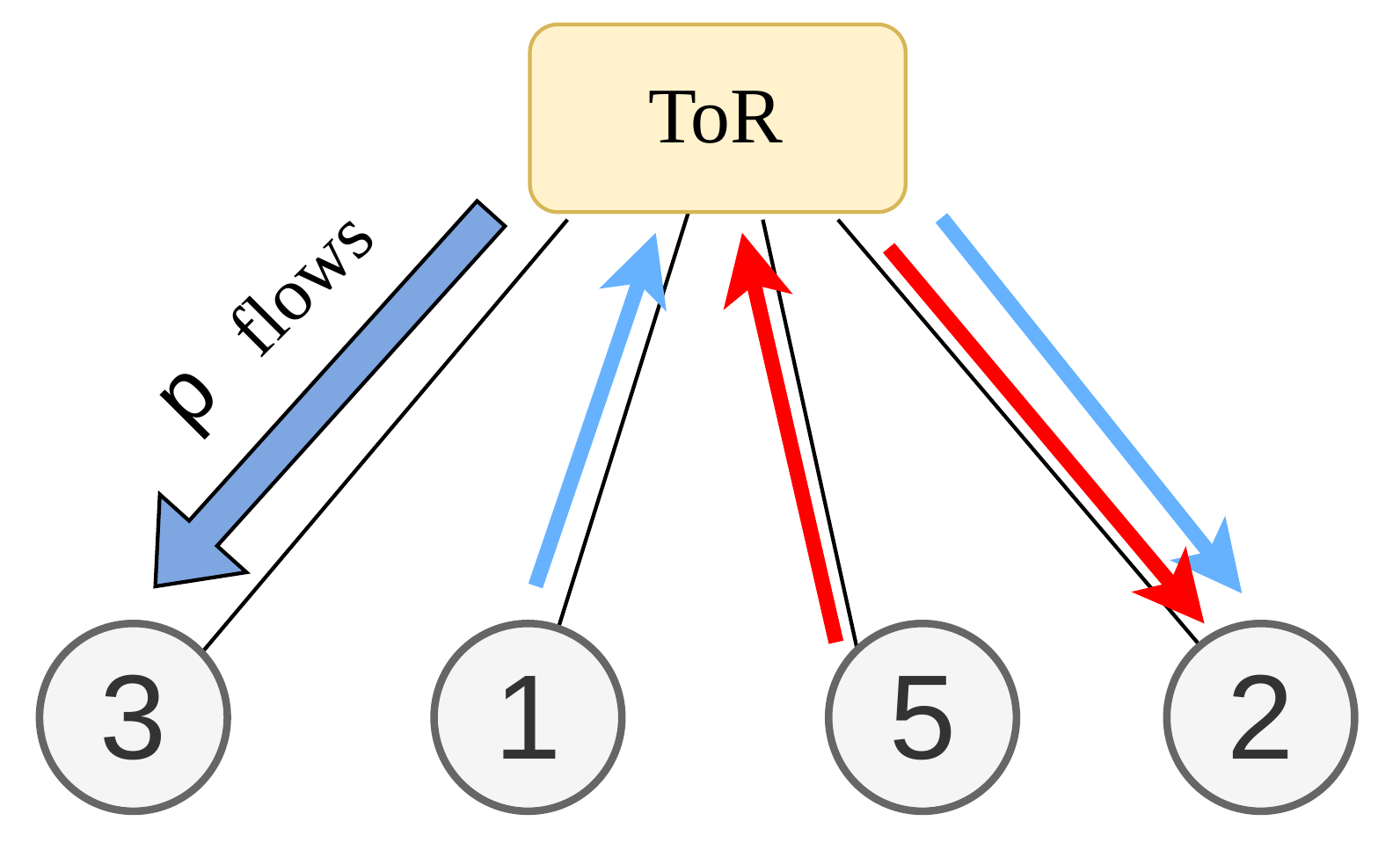}
            \caption{After replacement}
            \label{subfig:afterreplacement}
        \end{subfigure}
        
        \vspace{0pt}
        \end{minipage}
    \vspace{-10pt}
    \caption{Example showing benefit of swapping the congested node $3$ with $2$ in NCCL Tree-Allreduce algorithm. (a) and (b) show the logical flows between nodes in Tree-Allreduce. (c) and (d) show how the congested physical link is avoided by interchanging positions in the logical tree algorithm (the red and blue arrows are the chunks sent from $1$ and $5$ to $3$ before and to $2$ after replacement).}
    \label{fig:example}
    \vspace{-0.6cm}
\end{figure*}


We highlight the potential of our approach through an illustrative example. Our example considers the Tree-based AllReduce collective operation for a 16 node distributed training job. We consider the \collsched generated by NCCL as the starting point, that \sysname adapts in response to congestion. NCCL constructs two binary trees over the 16 nodes (referred as ranks 0-15), and divides each message into two chunks to be separately reduced using the two trees.
Fig~\ref{subfig:eventree} and \ref{subfig:oddtree} show the two binary AllReduce trees constructed using NCCL's heuristics. The trees are constructed such that the leaf nodes in one tree all have even ranks (Fig~\ref{subfig:eventree}), while those in the other tree all have odd ranks (Fig~\ref{subfig:oddtree}). The directed edges in the tree show the flows to be completed for the all-reduce: once each node receives all chunks from lower layers, it aggregates them before forwarding the result to the higher layer node. The reduce part of the collective is completed when each of the chunks has been aggregated at their respective root nodes (rank 15 and 0). In this example, it takes four time steps to complete it (which correspond to the four layers in the tree). The broadcast phase is started after this by sending the information back along the reverse edges in the two trees. 

For simplicity of exposition, our example assumes all 16 nodes are connected to the same switch (or ToR). Fig~\ref{subfig:beforereplacement} shows a subset of these nodes (other nodes not shown for brevity). Suppose the link propagation delay is $\alpha$  and the link capacity is $\frac{1}{\beta}$. 

\paragraphb{Performance without congestion} We first consider the scenario where there is no external congestion, and compute the time taken for the reduce operation in terms of $\alpha$ and $\beta$~\cite{HOCKNEY1994389}. Assuming the capacity at each link is fairly divided among the flows incident at the link, the total cost to reduce a message of size $s$ will be: $4\alpha + \frac{7}{2}\beta \cdot s$. (Explanation provided in the footnote.)~\footnote{The latency term comes from the 4 steps and bandwidth term for each of the steps are summed together: $(2+2+2+1)\beta\cdot\frac{s}{2}$ as bandwidth is shared on first three steps and message size is divided by two for each chunk.} 

\paragraphb{Congestion degrades performance} 
We next consider the case where there are $p$ external flows incident on node 3, reducing the fair share of flows incoming at node 3. This delays the data aggregation step at node 3, and the delay cascades over to higher layers (as node 7 must wait for input from node 3 before doing its aggregation, and so on). 
The cost for the reduce phase now will be $4\alpha+(3+\frac{p}{2})\beta\cdot s$. (Explanation provided in the footnote.)\footnote{Similar to previous calculation: $4\alpha+(2+(p+2)+1+1)\beta\cdot\frac{s}{2}$.} 

\paragraphb{\sysname adapts the \collsched to avoid congestion} Once \sysname detects incoming congestion at node 3 (as detailed in \S\ref{subsec:detect-congestion}), it attempts to swap nodes such that congestion is alleviated. Specifically, node 3 has two incoming flows in the current \collsched -- \sysname would find a swap that would reduce the number of incoming flows at the congested node 3, while still retaining the semantic correctness of the AllReduce operation. In Fig~\ref{subfig:eventree}, the only ``viable" replacements for node 3 are the green nodes. Interchanging with the yellow nodes will not help alleviate congestion as even after replacement, node 3 will still receive two incoming flows. The red nodes 4 and 12 are not viable because e.g., if 3 and 4 are interchanged in the even leaf-node tree (Fig~\ref{subfig:eventree}), then it is highly likely that the incoming flows to node 4 from nodes 1 and 5 in even leaf-node tree (Fig~\ref{subfig:eventree}), and 2 and 6 from the odd leaf-node tree (Fig~\ref{subfig:oddtree}) will compete for link capacity. We discuss in \S\ref{sec:design} how the set of viable nodes can be identified more generally for a given \collsched.  

In this case, we select node 2 randomly from the viable set of green nodes to be swapped with node 3.
This allows us to reach a configuration as displayed in Fig~\ref{subfig:afterreplacement} 
where the extra cost due to congested flows can be avoided. Note that we do not modify the odd leaf-node tree in this case as the congested node $3$ is already a leaf node there. This approach can be extended to handle more congested nodes by repositioning them on their corresponding trees. We can make similar optimizations on the reverse broadcast path for outgoing congested flows.

For a typical $\alpha = 4\mu s$, $\beta=1/(100\text{Gbps})$, $s=2\text{MB}$, and $p=2$, the time to complete the reduce phase increases from $576\mu s$ under no congestion to $656\mu s$ due to congested flows. The swapping of nodes helps in reverting the reduce phase time back to $576\mu s$ despite congestion-- a $14\%$ improvement (stronger performance wins can be seen in \S\ref{sec:eval} as we vary the settings, e.g. the number of participating nodes, the degree of congestion, the position of congested node in the pattern, etc). 




%% file: design-v2.tex
\section{Design}
\label{sec:design}

We describe how \sysname detects congestion at runtime in \S\ref{subsec:detect-congestion}. In \S\ref{subsec:safe-transform}, we discuss how \sysname pre-computes the set of safe transforms (viable swaps). 
Finally, in \S\ref{subsec:runtime-react}, we describe \sysname's runtime feedback loop. 


\subsection{Preliminaries and Notations}
\label{subsec:design-prelimnaries}


We represent a \collsched (\cs) as a time-expanded network (TEN)~\cite{tenFlowDep2002,BELIERES2021102203,ShuttleDispatching2021}, where each vertex $(t,r)$ corresponds to a rank (the logical id assigned to a GPU/compute node), $r$, at a logical time step, $t$, and each directed edge ($(t,r)\rightarrow(t+1,r')$) corresponds to a flow executed between two ranks in consecutive steps.

Fig~\ref{fig:unrolled dual tree} shows the complete unrolled TEN graph of NCCL's Dual Binary-Tree AllReduce discussed in the example in \S\ref{sec:overview} (this differs from Fig~\ref{fig:example} in that it also includes the broadcast phase of AllReduce). We additionally discuss Ring AllReduce and Recursive Doubling AllGather and their respective TEN graphs in Appendix~\ref{sec:nccl-collectives} (Fig~\ref{subfig:ten-ring} and Fig~\ref{subfig:ten-rc}).
We focus on these three collective patterns as our case-studies. Nonetheless, \sysname is a general system that can be used to tune other collective patterns.





\subsection{Detecting congestion}
\label{subsec:detect-congestion}

After running a batch of $B$ epochs, \sysname collects the flow completion times (FCTs) of all flows $F$ in the \cs\xspace in each epoch. Let $\mathbf{b}_f$ be the vector of the observed throughput (calculated using FCTs) in the most recent batch of $B$ epochs for a particular flow $f$. 
We now discuss how \sysname uses the TEN digraph $G$ and throughput vector of all flows $[\mathbf{b}_f]$ to detect which flows experience congestion, and the type of congestion. We differentiate between two types of congestion:




\paragraphb{(1) Steady congestion}
This type of congestion is persistently experienced by a flow in each epoch. It can arise when an unavoidable link (with no other multipath alternative) experiences congestion. For example, the link between a server and the next hop switch (typically, the top-of-the-rack or ToR switch) can face congestion due to multiple other flows running on the same server in parallel with the training job. These could be the storage traffic for checkpointing, data copy, and other such CPU-centric workloads~\cite{msft-gao}, or other \ML jobs scheduled on the same server (e.g. due to fragmentation of resources~\cite{crux,themis}, GPU sharing~\cite{siriusGPUsharing}, etc).\footnote{Chances of congestion on server-to-ToR links are lower in high-end deployments with over-provisioned servers that have dedicated NICs per GPU. Our work generally targets a wide-range of settings (including academic clusters and other low cost deployments) where the same NIC could be shared across GPUs and by storage traffic, leading to potential congestion in server-to-ToR links.}
Steady congestion can also arise if the spine layer of the underlying network topology is so heavily congested, that all paths through the spine layer experience performance degradation~\cite{congPatternsMetaIMC2025}.

Steady congestion is characterized by high FCTs for a flow in all epochs. We therefore identify whether a flow $f$ is \emph{steady} congested by checking
\vspace{-10pt}

\begin{equation}
\label{eq:server-cong}
    \texttt{p90}(\mathbf{b}_f)<\texttt{mean}(\mathbf{b}_{F(k_f)})(1-\delta_1)
\end{equation}

\vspace{-6pt}
where $\texttt{mean}(\mathbf{b}_{F(k_f)})$ is the mean throughput of all observed flows that have the same degree $k_f$ as of flow $f$ in the TEN graph (specifically, we consider the maximum of in-degree and out-degree). 
The $\texttt{p90}(\mathbf{b}_f)$ is the 90th percentile value of the observed throughput for flow $f$. Eq.~\ref{eq:server-cong} checks whether flow $f$ performs worse than other similar flows in the \cs\ for majority of the epochs. For our experiments, we set $\delta_1$ to $0.2$ as we expect a performance drop of at least 20\% under congestion for it to warrant attention. This threshold also helps avoid noise due to changes in the underlying stack and can be finetuned based on deployment. 

\paragraphb{\Irregular congestion}
Congestion also commonly arises in other links within the network that may not always lie on the flow's path in each epoch due to other multipath alternatives. For example, congestion arises in the spine-to-ToR links in a typical fat-tree topology in large-scale training jobs spanning multiple server racks~\cite{meta-rdma, alibaba-hpn, ncclx-meta}. Due to non-deterministic behavior of the underlying multi-path routing strategies (e.g. ECMP or adaptive routing), a flow may encounter a congested link on its path in one epoch, but not in another. We characterize the resulting congestion, observed only in a subset of epochs in the batch, as \irregular congestion. 
We identify whether a flow is \irregular congested by checking
\vspace{-10pt}

\begin{equation}
\label{eq:spine-cong}
    \texttt{normalizedSTD}(\mathbf{b}_f)>\delta_2
\end{equation}

\vspace{-6pt}
where $\texttt{normalizedSTD}(\mathbf{b}_f)$ denotes the normalized standard deviation of observed throughput. 
We set $\delta_2$=0.15 in our experiments.

Outside of congestion, another form of network performance degradation can arise due to partial link failures, e.g. where performance of a link drops from say 100Gbps to 30Gbps due to a switch dropping packets at a particular rate~\cite{flock,reps}. \sysname naturally extends to such failures too, treating them similar to a congested link. Such network failures are characterized as steady congestion when they arise in unavoidable server-ToR links, and as \irregular congestion otherwise.

        


\subsection{Finding Safe Transformations}
\label{subsec:safe-transform}

Next, we describe the transforms-- the replacement strategies-- used by \sysname to avoid congested flows. We start with the TEN digraph for the given \cs\xspace and detect congested flows. Once we detect a congested flow, we mark its source and destination as problem nodes. We want to avoid traversing this flow in the collective graph in subsequent epochs. To resolve congestion, we select one of the problem nodes and swap it with other nodes in the TEN graph. We now discuss how these swappable nodes are selected.


\subsubsection{Finding semantically viable swaps}

We want to swap nodes while ensuring that the new pattern (represented by its TEN graph) retains the semantics of information exchange for the collective operation.
We discuss three swapping transforms that ensure this. 


\noindent{\textcircled{1}} \textbf{Global Permutation swap:} 
The main idea is derived from the observation that most popular "All" collective operations (AllReduce, AllGather etc) are equivalent under permutation of ranks, in that we can swap all instances of a device $x$ with all instances of device $y$ in the TEN graph, and 
the pattern and still remain semantically equivalent. Note that this transformation is similar to finding an appropriate permutation of rank to device mapping. 

\begin{figure}
    \centering
    \begin{subfigure}{.45\linewidth}
        \includegraphics[width=\linewidth]{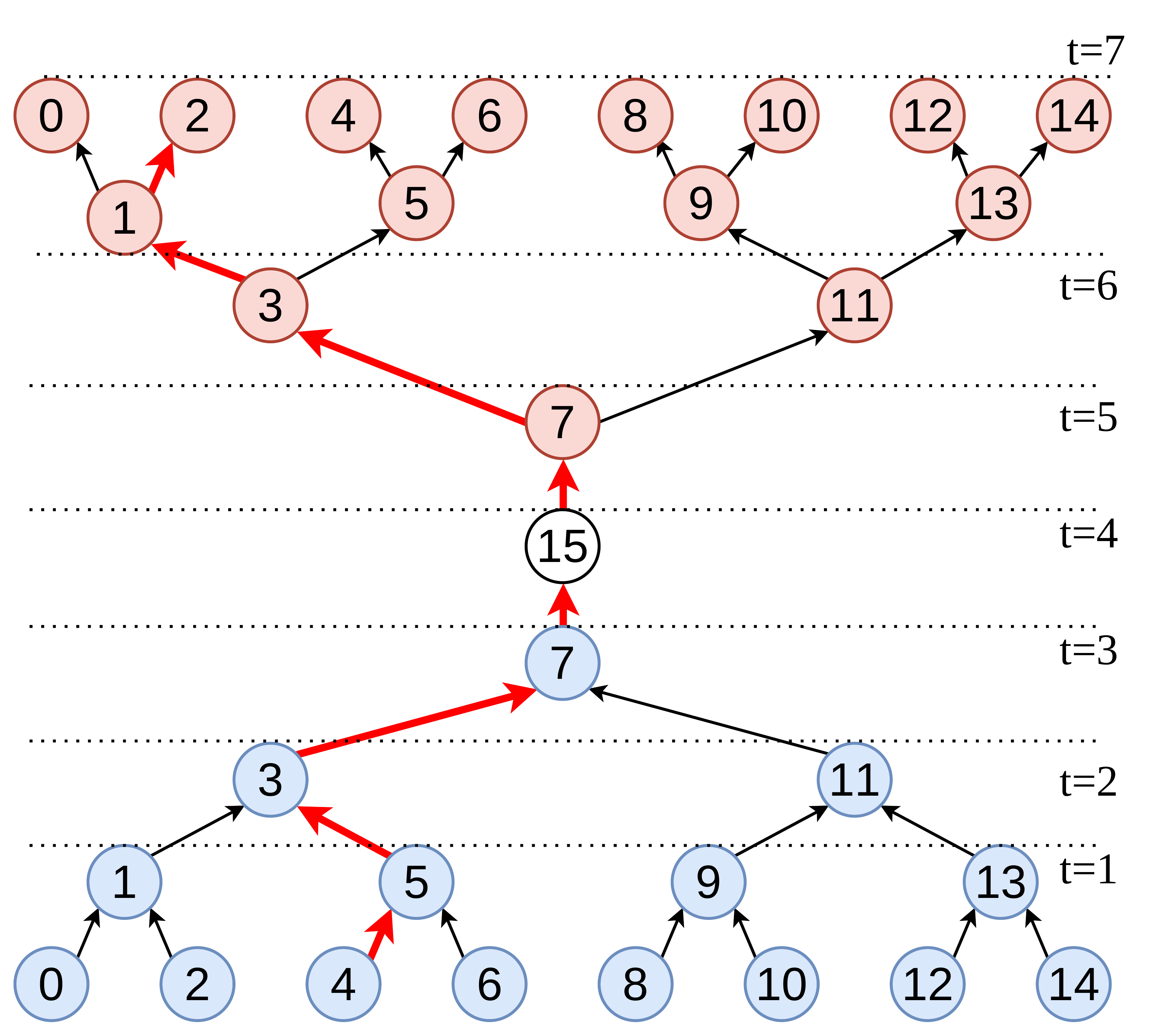}
        \caption{Even-leaf tree}
        \label{subfig:unrolled-eventree}
    \end{subfigure}%
    \begin{subfigure}{.45\linewidth}
        \includegraphics[width=\linewidth]{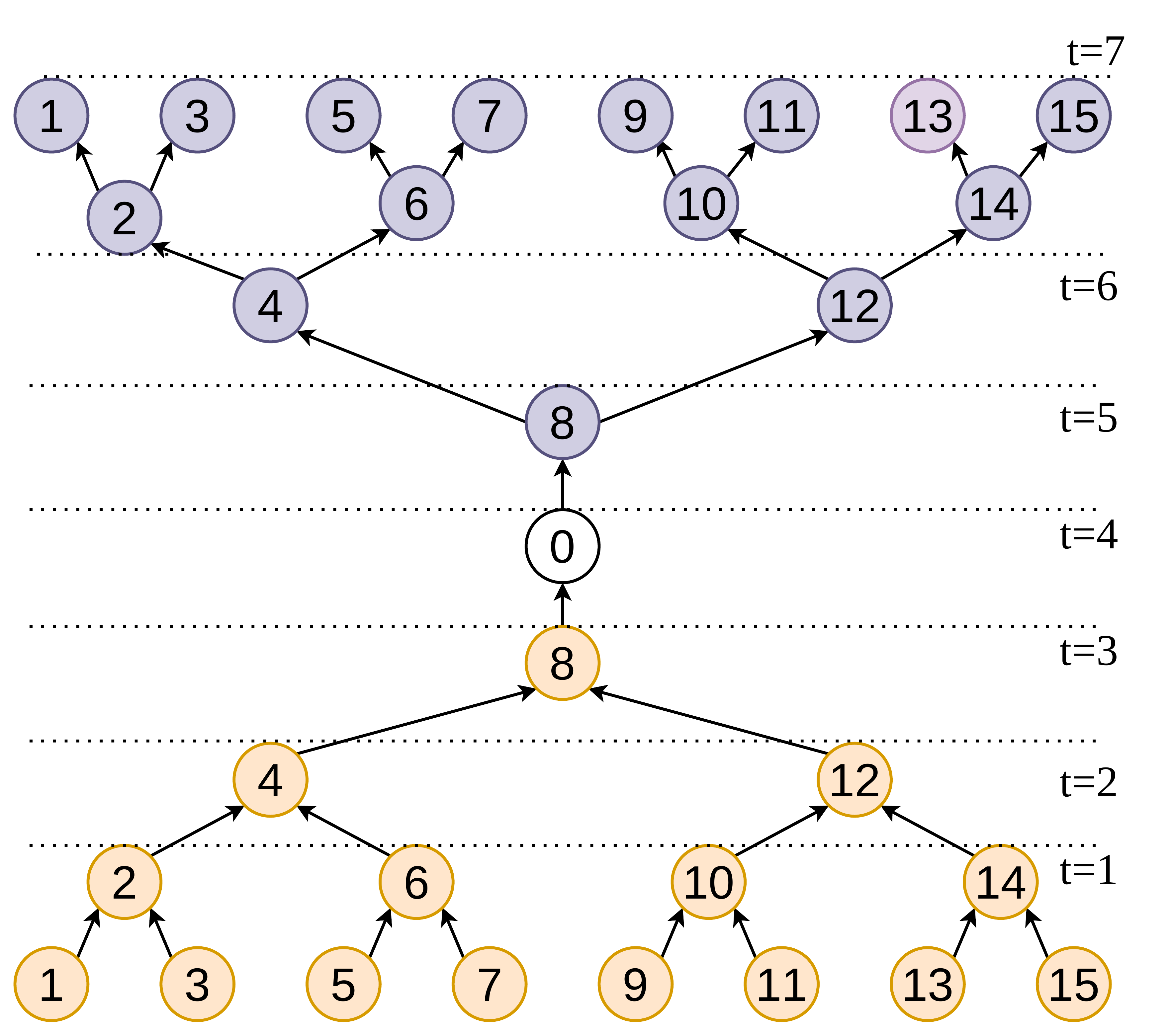}
        \caption{Odd-leaf tree}
        \label{subfig:unrolled-oddtree}
    \end{subfigure}
    \vspace{-10pt}
    \caption{TEN of NCCL dual-tree AllReduce}
    \vspace{-0.7cm}
    \label{fig:unrolled dual tree}
\end{figure}

We can apply this swap in a dual tree transform (Fig~\ref{fig:example}) by replacing all instances of congested node 3 with all instances of node 2 in both trees (Fig~\ref{subfig:eventree} and \ref{subfig:oddtree}). If we face \irregular congestion (spine-TOR links) in a flow of the dual-tree collective, this transform helps avoid that particular spine link by changing source/destination. However, if the congestion is in the server-TOR links, the performance deterioration persists as the overall degree of congested node 3 remains same before and after the swap, across the collective. Note that the transforms have different effects depending on the initial \cs-- we discuss {~\textcircled{1}} in the context of other collectives in \S\ref{subsubsec:applying-transforms}.




\noindent{\textcircled{2}} \textbf{Chunk-local Permutation Swap:}
Typical \cs s divide the message to be transmitted into chunks. Each chunk is transmitted independent of the other chunks, e.g. the even-leaf and odd-leaf trees in Fig~\ref{fig:example} transmit two different chunks separately. We identify independent chunk graphs in a \cs\xspace and apply the \emph{Permutation Swap} transform selectively on only the affected chunk graphs. Fig~\ref{fig:unrolled dual tree} shows the unrolled TEN digraph for the dual-tree AllReduce algorithm, where we can swap nodes in both trees independently. E.g., if node 15 in Fig~\ref{subfig:unrolled-eventree} faces congestion, its effects might be alleviated if all of its instances in Fig~\ref{subfig:unrolled-eventree} are swapped all instances of an uncongested node 10, without modifying any nodes in Fig~\ref{subfig:unrolled-oddtree} digraph.\footnote{The swap can help alleviate \irregular congestion between node 7 and 15. In the case of steady congestion, it removes 15 from being the bottleneck of the pattern and shifts it to a location where it is less likely to affect.}

It is more desirable to use such a finer grained swap (as compared to global permutation) when possible, as it allows making selective changes, local to wherever performance deterioration is detected. Making more changes than necessary can add to the overhead in terms of exploring more unused network paths.


\noindent{\textcircled{3}} \textbf{Position-equivalent Swaps:}
The above two transforms are effectively permutation of rank to device mapping applied on the \collsched and chunk-graph scale respectively. Next, we apply a brute force algorithm (formally defined in Appendix~\ref{subsec:appendix:brute-force-analysis}) on each chunk graph to identify the equivalent positions that can be swapped while maintaining semantic equivalence of the \cs. The algorithm considers all pairwise positions and checks whether swapping them violates the original collective operation.
We check this by assigning all nodes a bitvector with a single bit set to 1 based on rank. We simulate an execution of the collective operation on this vector by running a breadth-first traversal of the collective graph and checking whether the final bitvector on all nodes aligns with the conditions of the collective operation. For the popular collective patterns, the brute force algorithm has an $O((N\log N)^3)$ complexity to find the chunk graphs and equivalent sets of swappable nodes. We only generate this once at the start of the job and can refer to it at runtime.

Fig~\ref{fig:unrolled dual tree} shows the equivalent positions in dual-tree AllReduce algorithm where the nodes with same color form an equivalent set and can be freely swapped.

Note that the search space of transform {~\textcircled{3}} is different than the previous two transforms. This transform allows even finer grained changes than~{~\textcircled{1}} and~{~\textcircled{2}} that help make precise decisions in resolving congestion. Eg, if some  node is steady congested, we only want to swap its placement wherever it lies on the critical path. Changing other positions may not help improve the collective completion time but can add additional uncertainty. 
Based on this granularity of changes, for a given congested node, we prioritize available~{~\textcircled{3}} transforms over~{~\textcircled{2}} and~{~\textcircled{1}}.


\subsubsection{Pruning the Swap Space}
\label{subsec:transforms-common-remarks}

Given a congested node, we get a set of transforms from above that can be applied to possibly alleviate congestion.
We further prune this set by checking that after applying the transform, for the complete \cs, the degree of each rank at various timesteps does not exceed the degree of the original rank (before the swap) in this position.
In the example in Fig~\ref{fig:example}, this check removes all the red nodes from the set of viable nodes. For the given congested node, this pruned set is the set of viable transforms we have available (yellow and green nodes in Fig~\ref{subfig:eventree}).
Among the set of viable transforms, we further prioritize swaps that reduce the degree of congested nodes (the green colored nodes in Fig~\ref{fig:example}).
We randomly choose a node to swap from this prioritized set (e.g. node 2 in Fig~\ref{fig:example}). If none of the prioritized nodes help improve performance (as discussed in \S\ref{subsec:runtime-react}), we consider transforms with other viable nodes (i.e. the yellow nodes in Fig~\ref{fig:example}).  

\subsubsection{Applying transforms on collectives}
\label{subsubsec:applying-transforms}


\begin{table}[]
    \centering
    \begin{tabular}{|c|c|c|c|}
    \hline
        \textbf{\small{Transforms}} & \textbf{\small{Ring}} & \textbf{\small{Recursive D.}} & \textbf{\small{Tree}} \\\hline
        \textcircled{1} & \small{Spine, Failure} & \small{All} & \small{Spine, Failure} \\ 
        \textcircled{2} & \ding{56} & \ding{56} & \small{All} \\ 
        \textcircled{3} & \ding{56} & \small{Spine, Failure} & \small{All} \\\hline
    \end{tabular}
    \vspace{-9pt}
    \caption{Comparing the types of congestion (\S\ref{subsec:detect-congestion}) resolved by the transforms for popular CAs (Ring-based algorithms, Recursive Doubling AllGather, Tree based algorithms). \emph{Spine} is Spine-ToR congestion, \emph{Failure} is network failures, \emph{All} is all congestion types in \S\ref{subsec:detect-congestion}. (\ding{56}) marks that the transform is not able to help with the \cs, topology and congestion scenarios we consider.}
    \label{tab:transforms-summary}
    \vspace{-15pt}
\end{table}

While we exemplify the above transforms in the context of AllReduce trees, they are broadly applicable. We also apply and evaluate these transforms for Ring AllReduce and Recursive Doubling AllGather (these {\collsched}s have been explained in Appendix~\ref{sec:nccl-collectives}). 
The applicability of REACT's transforms depends on the structure of the underlying collective and the type of congestion.
Table~\ref{tab:transforms-summary} summarizes the applicability of the transforms to the three transforms we study and the different congestion scenarios. 
Tree AllReduce supports all three transforms; Ring AllReduce primarily benefits from global permutation, since changing node positions within individual chunks does not reduce node degree; and Recursive Doubling supports global permutation and position-equivalent swaps. 
We provide detailed examples and analysis for each collective in Appendix~\ref{sec:appendix:ring-rc-transforms}.
Other popular collectives like ReduceScatter, Broadcast, Reduce, Scatter etc. have similar {\collsched}s and our work can be applied for them as well. 

\subsection{Runtime Feedback Loop}
\label{subsec:runtime-react}

\begin{figure}
    \centering
    \includegraphics[width=0.9\linewidth]{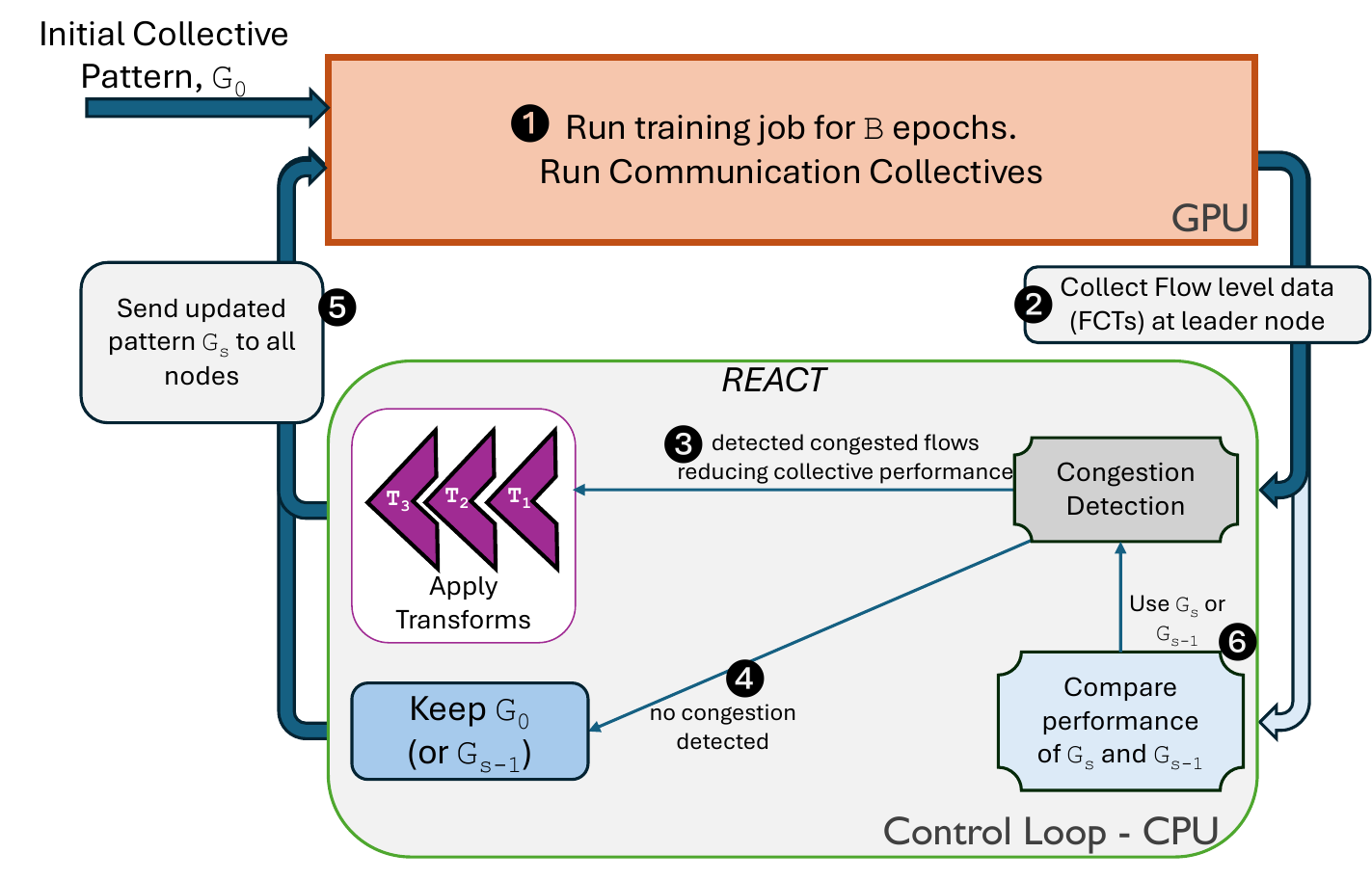}
    \vspace{-9pt}
    \caption{State diagram of \sysname. The orange box marks the distributed training workload running on the GPUs that uses the input collective pattern for communication. The gray boxes run on a separate thread on the CPU-- collecting flow level data from all nodes, running the control loop, finding a new collective pattern, and sharing it with the training job processes on all nodes.}
    \label{fig:ct-state-diagram}
    \vspace{-15pt}
\end{figure}

Fig~\ref{fig:ct-state-diagram} shows the state diagram of \sysname's runtime feedback loop. 
At the start of the job, we take the initial collective graph $\texttt{G}_0$.
We run the training job with $\texttt{G}_0$ and once a batch of $B$ epochs is completed (\ding{182}), all flow completion times (FCTs) in the \cs\xspace are collected at the leader node (\ding{183}). We discuss more on how the collection and sharing of FCTs is implemented in \S\ref{sec:impl}.
Next, we identify the congested flows using methods from \S\ref{subsec:detect-congestion}. 
If no congestion is detected (\ding{185}), the control loop decides to keep the current \cs\xspace graph.
If congestion is detected (\ding{184}), we proceed to applying viable transforms on $\texttt{G}_0$. 
The updated graph, $\texttt{G}_1$, is shared with all nodes by the leader node (\ding{186}). We explain these steps in \S\ref{subsubsec:replace-nodes-loop}. \sysname's feedback loop then repeats these steps, taking the updated graph $\texttt{G}_1$ as input to produce the next graph $\texttt{G}_2$, and so on (as detailed in \S\ref{subsubsec:revert-state-loop}). 

\subsubsection{After a single batch of epochs} 
\label{subsubsec:replace-nodes-loop}

We now discuss how, after running a batch of $B$ epochs with an input pattern $\texttt{G}_0$, we apply transforms to $\texttt{G}_0$ and swap out congested nodes to get an updated pattern $\texttt{G}_1$. The algorithm pseudo code can be found in Appendix~\ref{sec:appendix:transforms}.

We begin with allocating weights to each edge in the input TEN graph, $\texttt{G}_0$, based on completion time of each flow in each epoch in $B$. We then find the critical paths in each of these weighted graphs, where the critical path is the one with the highest edge weights (total completion time).
We also construct another graph where the edge weights are derived from the average FCTs across all epochs in batch $B$, and compute its critical path. 
Reducing the time taken along the critical path will help reduce the overall collective communication time. Therefore, we only consider swapping out congested nodes that lie on at least one of these critical paths. Note that we construct critical paths for each epoch individually (along with looking at the average FCTs across epochs) because \irregular congestion would impact only a subset of epochs. 

We want to prioritize resolving \emph{steady} congestion as it is more severe than \emph{\irregular} congestion.
We can further identify whether a node is steady congested by checking if all incoming/outgoing flows to a node in the graph are \emph{steady} congested. 
Thus, we sort all vertices $x=(t,r)$ in $\texttt{G}_0$ based on the number of congested flows that have the vertex as a source or destination. We iterate over the list and for each congested node, find a viable list of  transforms  (as detailed in \S\ref{subsec:safe-transform}).

If multiple transforms with same priority are available, we randomly select a replacement for the congested node among the viable nodes (\S\ref{subsec:transforms-common-remarks}), to get a new collective pattern, $\texttt{G}_0'$. We also update the initial $(B+1)$ critical paths (average and per-epoch in $B$) by applying the transform on each of them as well. We then estimate the performance improvement offered by this transform using a simple analytical model. We estimate the FCTs (i.e. the weight of each edge) in $\texttt{G}_0'$ based on mean observed FCTs for the corresponding source-destination pairs over the previous batches of epochs, scaled as per the new node degrees and message sizes. In case a transform introduces flows for which we do not have past data, we estimate its performance as the median completion time across all flows of same degree. We only accept the updated graph for the next iteration if the mean of path times of updated critical paths is at least $(1+\epsilon)$ times lower than the original.\footnote{Note that we compare the performance using old critical paths and choose not to recompute critical paths of $\texttt{G}_0'$ (for $\epsilon$-check) as it is possible for a TEN graph $\texttt{G}_0$ to have multiple critical paths of similar completion times -- we fix them iteratively.} In our experiments, we set $\epsilon$ to 5\%. 

Note that once we have resolved the \emph{steady} congestion with a transform above, we remove these steady congested nodes from the list of viable nodes for 
other congested nodes, so that they may not reintroduce congestion in future transforms. \Irregular congested nodes can still be reused, provided they pass the $\epsilon$ check above.

After each batch of epochs, we try to apply a maximum of $K_t$ transforms to ensure we explore for a bounded amount of time.
We set $K_t$ to 15 in our experiments but increase it every batch of epochs so that the search space expands in later epochs.
We obtain the updated graph $\texttt{G}_1$, after applying this set of transforms.  


\subsubsection{Over multiple batches of epochs} 
\label{subsubsec:revert-state-loop}
\vspace{-3pt}

Once a new graph has been generated after the above process, we run the next batch of epochs with the modified pattern $\texttt{G}_1$ (\ding{182}).
After the completion of the new batch of epochs, \sysname compares the the performance of the current modified \cs\xspace, $\texttt{G}_1$,  with the previous one, $\texttt{G}_0$ (\ding{187}). If the observed epoch communication time with $\texttt{G}_1$ is not less than $(1+\tau_r)$ of the previous batch of epochs with $\texttt{G}_0$, then \sysname reverts to the older \cs\xspace $\texttt{G}_0$ and tries new transforms on it, following the algorithm in \S\ref{subsubsec:replace-nodes-loop}. Else, we try to resolve congestion in $\texttt{G}_1$, if any, following the steps in \S\ref{subsubsec:replace-nodes-loop}. We set $\tau_r$ as 0.05 for our experiments as we expect each modification to provide at least 5\% improvement. Note that this check is based on the observed empirical performance, and differs from the check made for each transform in \S\ref{subsubsec:replace-nodes-loop} that was based on estimated (analytical) performance. 

We use this process to continuously explore for newer \collscheds and reach a more desirable solution (alleviating as much congestion as possible). 
It is possible that we may have reached a good enough solution and not want to explore any further. Hence, we set $E_{max}$, to 5, as the number of exploratory rounds (where each round has $B$ epochs), after which we stick to the current \cs\xspace if no better pattern was found. We set the batch of epochs size, $B$ to be 6 in our experiments. Moreover, if no exploration was done over $E_{max}$ rounds (i.e. $E_{max}B=30$ epochs), we trigger exploration in the next round, to ensure we are not stuck in a local optima.


Since a cluster can have varying jobs and network performance, we delete all old FCT-data from beyond the prior $H$ epochs. This helps account for any job churn in the cluster and we set it to 60 epochs for our experiments.

%% file: implementation-v2.tex
\section{Implementation}
\label{sec:impl}


\begin{figure}
    \centering
    \includegraphics[width=0.8\linewidth]{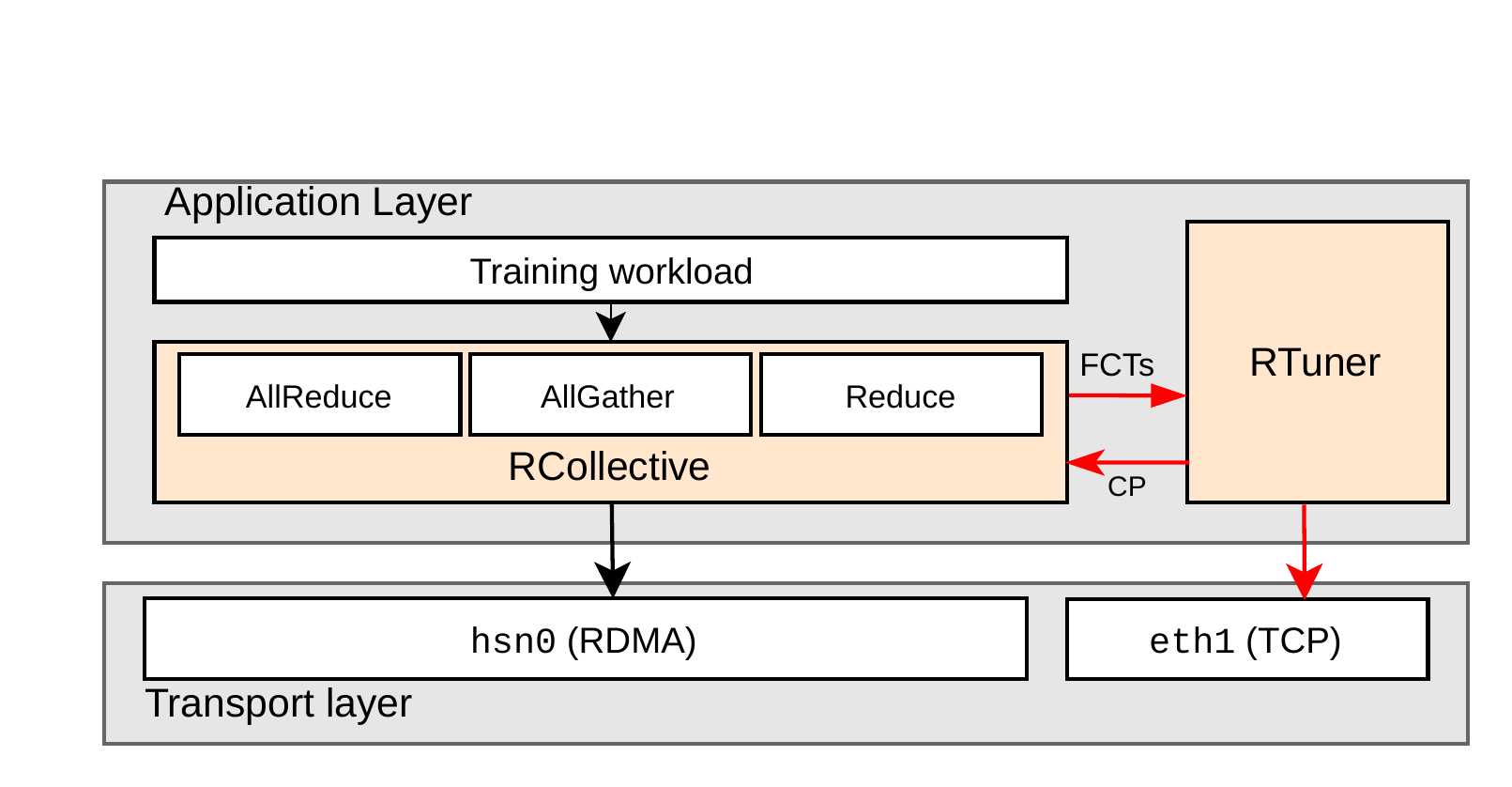}
    \vspace{-13pt}
    \caption{\sysname architecture.
    }
    \label{fig:impl}
    \vspace{-23pt}
\end{figure}



We implement REACT at the application layer with two components: a collective execution library (RCollective) and a runtime tuner (RTuner). Figure~\ref{fig:impl} shows the architecture of REACT. 
We implement the collectives such as AllReduce, AllGather etc in the collective library.
We implement the tuner to collect flow-completion statistics across nodes, compute an updated collective pattern, and disseminate it to all participating nodes.

\subsection{RCollective library}
\vspace{-3pt}

We implement the RCollective library as a wrapper using the peer-to-peer (p2p) communication APIs-- \texttt{isend} and \texttt{irecv} in PyTorch. Given a TEN graph of a communication collective, chunk sizes and specific intermediate aggregation steps, we implement RCollective to post corresponding communication kernels to the GPU asynchronously. We use separate sets of streams on the GPU for each chunk sub-graph of the collective algorithm so that maximum operations can run in parallel. 
We also ensure the data dependency between communication steps by adding blocking commands between the streams within a chunk sub-graph. We post all the kernels at the start of the collective and rely on the GPU kernel scheduler for efficiency. See Appendix~\ref{subsec:ctcoll-algo} (Algorithm~\ref{algo:ctcollective}) for the detailed pseudo-code for implementing a custom collective pattern using p2p APIs. 


We also measure flow completion times by inserting timing events before and after each peer-to-peer operation. These events capture the elapsed time on the GPU stream until the transmitted data becomes available to the next dependent kernel.


\noindent
\textbf{Overheads.} The RCollective library lies on the main performance path of the training workloads. The overheads due to multiple streams and flow timing events are minimal as these objects are created once at the start of the job and reused throughout the job lifetime. Other overheads occur due to setting up of new NCCL ``communicator objects" when a new graph algorithm is sent by RTuner and missing out on the buffer management optimizations of NCCL. These overheads are discussed in more detail in \S\ref{sec:eval}.




\noindent
\textbf{Comparison with NCCL.}
A tighter integration with NCCL would likely reduce prototype overheads and provide a cleaner implementation path. In principle, NCCL's ext-net profiler interface for custom network support could be repurposed as a lightweight wrapper around the underlying transport, with hooks to measure flow start and completion times. In practice, this was difficult in our prototype deployment because the shared academic GPU cluster we use relies on the Cray Slingshot network ~\cite{cray}, whose transport stack is exposed through cluster-specific libraries (e.g., \texttt{libfabric.so}) and a NCCL \texttt{ext-net} plugin. This makes it difficult to insert our custom measurement hooks without modifying the underlying system software.

\subsection{REACT Tuner}

We configure the REACT Tuner (RTuner) component with one leader node and rest as workers. The leader is responsible for collecting all of the data from various devices, generating a new collective pattern based on the observed state, and sending the new graph to all other nodes. The worker processes send the measured flow completion times after every batch of $B$ epochs and wait to receive the new graph from the leader. 

We run RTuner on a separate CPU thread so as to minimize any interruptions to the main training process and it is initiated at the start of the job. The flow times and graphs are exchanged with RCollective library via shared memory. We run the leader RTuner process on rank 0 by default. We use the \texttt{gather} and \texttt{broadcast} communication collective operation in the Gloo~\cite{gloo} library to collect flow times from all nodes and send the new graph.


\noindent
\textbf{Overheads.} 
RTuner primarily consumes additional CPU resources. In our setting, this overhead is small relative to the overall cost of distributed GPU training, since the dominant runtime cost remains GPU computation and communication. RTuner also uses a different network interface from the data path as discussed in \S\ref{subsec:proto-shared-cloud}.
\vspace{-5pt}


\subsection{Prototype on a shared cloud} 
\label{subsec:proto-shared-cloud}

We run our system on a national academic GPU cluster. 
Each server in the cluster has 4 A100 GPUs connected via NVLink. The servers are all connected by a 200Gbps Cray Slingshot~\cite{cray} network \texttt{hsn} interface which is used by the RCollective to run communication. All nodes also support a Intel Corporation I350 Nic \texttt{eth1} interface which is used for cluster management and sending control messages.\footnote{Many GPU clusters support such a configuration with 2 NICs- one for fast data path and second for slower control path.} RTuner uses \texttt{eth1} as the interface to exchange control loop information via Gloo to avoid disturbing the collective performance. Further, to avoid making the main job wait for updates from RTuner, we let the main job continue for an additional 5 epochs with the older collective pattern. This allows RTuner to complete the consensus while the training job keeps running without any additional wait time. See Appendix~\ref{sec:impl-additional-details} (Fig~\ref{fig:topo-delta}) for the intra-node and the Slingshot network topology.

%% file: eval.tex
\section{Evaluation}
\label{sec:eval}
\vspace{-3pt}


We discuss our evaluation methodology in \S\ref{subsec:eval-method}. Then, in \S\ref{subsec:testbed}, we present our evaluation on a small-scale testbed implemented on a national academic shared cluster (as described in \S\ref{sec:impl}). Finally, in \S\ref{subsec:eval-sim}, we  present our larger-scale ns-3 simulation results under different scenarios. 

\vspace{-3pt}
\subsection{Evaluation Methodology}
\label{subsec:eval-method}
\vspace{-3pt}


\noindent\textbf{Collective Benchmark.} We run collective operations in a tight loop over multiple epochs on a pre-defined set of nodes. We experiment with different collective operations -- NCCL's tree AllReduce, Ring, Recursive Doubling. \radhika{and AdapCC's solver-generated collectives.} We evaluate how long the collective communication takes over epochs with and without \sysname under different congestion scenarios. 

\noindent\textbf{Modeling Congestion.} 
We model congestion in three ways. First, we inject external background flows from nodes that are not participating in the collective into one or more nodes that participate in the collective. We define the congestion degree $p$ as the number of external flows injected per affected source-destination pair. Second, we run two jobs simultaneously with a configurable number of overlapping nodes to model resource fragmentation and server-level contention. Third, we run two simultaneous jobs with no shared nodes but overlapping network paths to isolate network-only contention. The latter two are only evaluated in simulation, while the first congestion scenario is used both in our testbed and simulations. 

\noindent\textbf{Network Topology.} For the testbed evaluation, we use the provided Slingshot topology (See Apendix~\ref{sec:impl-additional-details}, Fig~\ref{subfig:network-topo}). 
We further evaluate \sysname in simulation on three network topologies: a star topology, a 128-server Clos topology~\cite{kfattree}, and a topology derived from a publicly available Alibaba trace~\cite{crux,alibaba-trace}.
The Alibaba topology has 48 TORs, each connected to the 3 aggregate switches. Each server is connected to exactly one TOR. We simulate a 720-server Alibaba topology. 

\noindent\textbf{Metrics.}
We measure the total collective completion time per epoch and also calculate the algorithm bandwidth by dividing the message size by the collective time, as done in prior works~\cite{taccl,teccl}. We compare the improvement in algorithm bandwidth (or reduction of collective completion time) of \sysname against the baseline under the various scenarios.


\vspace{-5pt}
\subsection{Testbed Evaluation}
\label{subsec:testbed}
\vspace{-5pt}

\begin{table}[t]
\centering
\begin{tabular}{|c| c| cc| c|}
\hline
\multirow{2}{*}{\#Nodes} & \multirow{2}{*}{$p$} & \multicolumn{2}{c|}{Speedup $\times$} & \multirow{2}{*}{\#$\overline{changes}$} \\
\cline{3-4}
 &  & Mean & p95 & \\
\hline
\multirow{3}{*}{4}
    & 0 & \textcolor{deepred}{0.964} & \textcolor{deepred}{0.999} & 34 \\
    & 1 & {1.138} & 1.101 & 35.6 \\
    & 2 & 1.348 & 1.145 & 33.2 \\
\hline
\multirow{3}{*}{8}
    & 0 & \textcolor{deepred}{0.963} & \textcolor{deepred}{0.980} & 41.25 \\
    & 1 & 1.180 & 1.165 & 40 \\
    & 2 & \textbf{1.383} & 1.222 & 36.25 \\
\hline
\end{tabular}
\vspace{-5pt}
\caption{Ratio of mean (or p95) collective completion time over epochs for baseline to \sysname, across node counts and congestion degree, $p$, for Tree AllReduce on GPU cluster deployment.}
\label{tab:perf-ratio}
\vspace{-15pt}
\end{table}

We first show the feasibility of our approach and its benefits by implementing \sysname on a shared cluster, as discussed in \S\ref{sec:impl}. 
We run two benchmark jobs with 4 and 8 A100 GPUs respectively (Table~\ref{tab:perf-ratio}). We enable adaptive routing
and GPU Direct RDMA in the underlying Cray Slingshot 200Gbps network~\cite{cray} for our experiments. We run the dual binary Tree AllReduce for 5k epochs with a 100MB allreduce message size. 

We request one GPU each on different nodes from the cluster scheduler to evaluate \sysname under inter-server network congestion. 
Since we cannot control the node allocation from the cluster scheduler in terms of node placement in the topology, we run our benchmarks multiple times to try different resource allocations. For fairness, we only make comparisons by running both the baseline and \sysname on the same allocation one after the other. Since node placement in the topology can affect the base latency and observed bandwidth, we measure the speedup by taking a ratio of the latency of baseline and \sysname for the respective placement. We measure the speedup for both the mean and p95 collective completion time over the epochs.

We introduce external flows for congestion in the cluster, from the CPU of an external node to the CPU of a rank running the training job. This external flow is an RDMA ping pong flow that periodically sends 1GB of data. We vary the number of external flows as $p=0,1,2$, to evaluate performance under no congestion, and different degrees of congestion.
Table~\ref{tab:perf-ratio} shows a 13-38\% speedup  offered by \sysname under congestion ($p=1,2$).

We see a $<4\%$ drop in performance for when we run the benchmark under no congestion scenario, compared to the baseline. 
This cost is due to two reasons-- overhead of exploration (when we see variable flow performance at runtime) and cost of setting up new connections-- without external congestion to be alleviated and compensate for the cost. We try to minimize this cost in our design and discuss in \S\ref{sec:discussion} on how to reduce this cost even further. \sysname is thus a feasible design that improves performance of communication tasks for distributed training workloads in a shared GPU cluster, with low overheads.

\vspace{-5pt}
\subsection{Simulations}
\label{subsec:eval-sim}
\vspace{-5pt}

We now evaluate \sysname under a variety of congestion scenarios (\S\ref{subsec:eval-method}) and under different network stacks. We use ns3~\cite{ns3} packet level simulator to simulate a collective communication job. We use DCTCP~\cite{dctcp} as the congestion control with Random early detection (RED) queues on switches to mark ECNs (queue thresholds set as 32 and 60 for low latency). We size the switch buffers at 32MB and prioritize ACKs in switch queues to prevent delays. For all experiments, we use ECMP as the underlying routing scheme. Following our discussion with the industry, we change ports every epoch to allow ECMP to not be stuck with a bad path choice. Since creating new connections is costly for each epoch, each flow in the \collsched is initialized with 4 connections on 4 different ports respectively. At each epoch, one port is chosen at random to send the data. We simulate with all links set to 100Gbps and link latency as $1\mu s$. Unless specified, for all experiments we simulate an 8-node training job for 100 epochs with a message size of 20MB. We only measure the mean algorithm bandwidth of a job after 25 epochs to allow initial epochs for \sysname to converge. 

\vspace{-5pt}
\subsubsection{Congestion due to varying number of external flows}
\label{subsubsec:tree-and-rc}
\vspace{-5pt}

\begin{figure}[t]
\centering
    \begin{subfigure}[b]{0.48\textwidth}
       \includegraphics[width=1\linewidth]{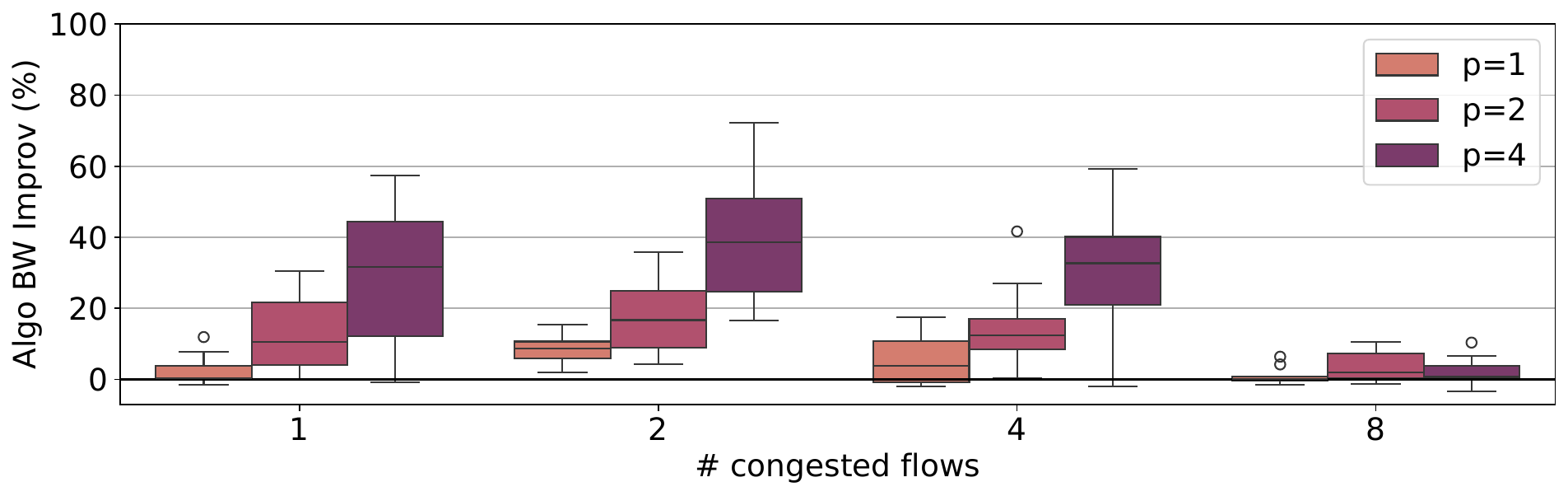}
       \vspace{-18pt}
       \caption{8 node Dual Tree AllReduce}
       \label{subfig:fattree-20mb}
       \vspace{-2pt}
    \end{subfigure}
    
    \begin{subfigure}[b]{0.48\textwidth}
       \includegraphics[width=1\linewidth]{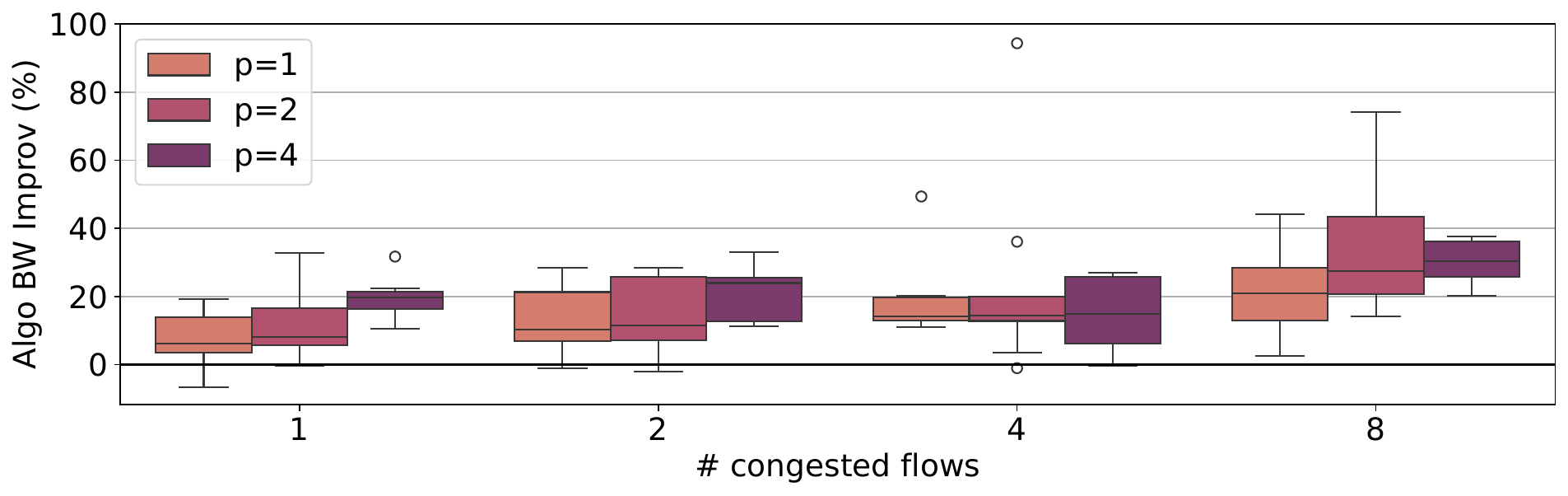}
    \vspace{-18pt}
    \caption{8 node Recursive Doubling AllGather}
    \label{subfig:recursive-doubling}
    \end{subfigure}
    \vspace{-22pt}
    \caption{Mean algorithm bandwidth improvement for a sweep of 10 randomly generated scenarios under various degrees of network congestion on a Fattree topology and 20MB message size. $x$-axis marks the number of source and destination pairs that initiate $p$ external flows each.}
    \label{fig:coll-bw}
    \vspace{-18pt}
\end{figure}

We simulate different collectives (Tree AllReduce and Recursive Doubling AllGather) under changing number and degrees of external congested flows on a 3-layer Clos topology. We find that the performance deterioration is different when different ranks in a \collsched are subjected to congestion from external flows. To make a more comprehensive comparison, for each case of number of congested flows (source-destination pairs of external flows, $x$-axes in Fig~\ref{fig:coll-bw}) and degree of congestion (number of flows between same source-destination pair, $p$), we randomly select the source (from nodes participating in collective) and destination ranks (remaining nodes in the topology) for background traffic. We randomly generate 10 scenarios for each case and plot the percentage improvement in the mean algorithm bandwidth over the epochs in Fig~\ref{fig:coll-bw}. We find that even in low congestion scenarios, $p=1$, \sysname can help in alleviating performance by upto 20\% for both collectives in the worst case scenario-- if the cluster is in a particularly bad configuration. In high congestion scenario ($p=2,4$), we find that \sysname can alleviate performance upto 60-95\%, with mean across scenarios being 35\% and 15\% for tree and recursive doubling respectively.

\noindent\emph{Note:} In the case of 8 congested flows, all nodes in the training job for tree AllReduce are congested allowing for no room for swaps. Similarly, Recursive Doubling benefits by reducing the path distance for later time steps (avoiding Spine-ToR congestion) when message sizes are large, and hence it is still able to be improved upon in high congestion scenarios.

\subsubsection{Tail Latency and effects of ECMP.}
\vspace{-4pt}

We further evaluate the tail epoch communication times, with and without \sysname for different message sizes, and the effects of varying the number of ECMP port choices (see Appendix~\ref{subsubsec:tail-and-ecmp}). We observe that exploiting multiple path choices through ECMP can help reduce congestion on network links as the number of ports increases, similar to the scenarios described in~\cite{meta-rdma}. However, even with more port choices, collective communication performance can still degrade in the presence of background flows, and \sysname consistently improves performance across different routing configurations.
Thus, \sysname also reduces the need to maintain higher number of ECMP ports or RDMA queue pair connections which reduces the overall memory load on the NICs to allow for other tasks.
We also compare performance of REACT for different message sizes in Appendix~\ref{subsubsec:tail-and-ecmp}.

\vspace{-2pt}
\subsubsection{Multiple Jobs.}
\vspace{-3pt}

\begin{figure}[t]
\centering
       \includegraphics[width=1\linewidth]{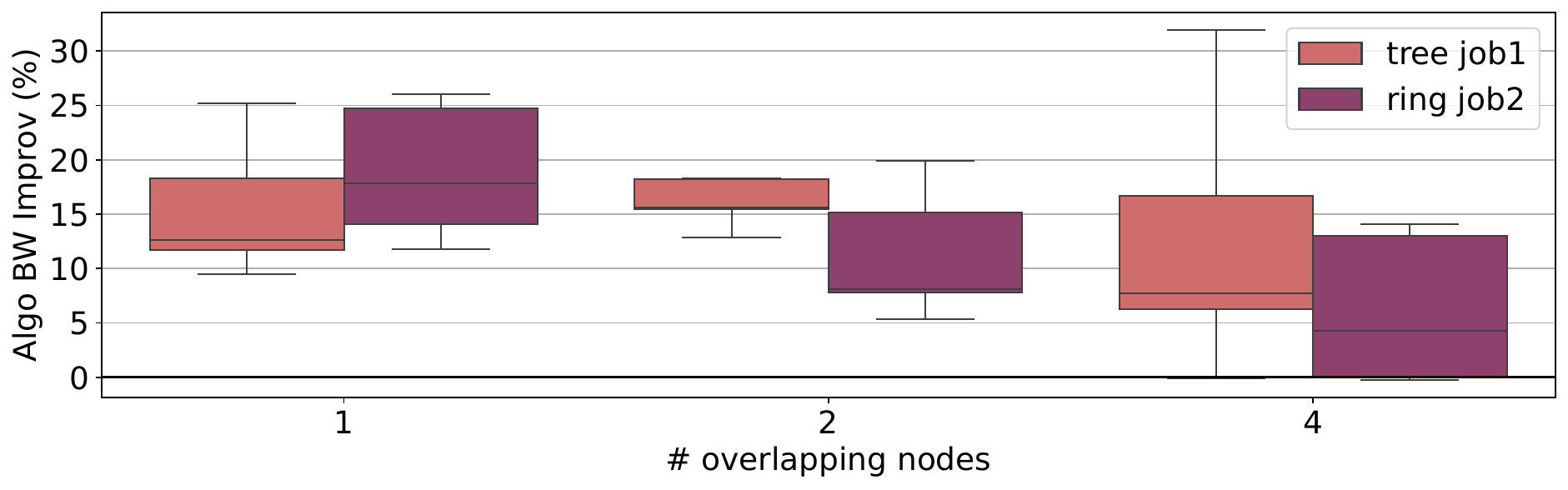}
    \vspace{-23pt}
    \caption{Percentage performance improvement with REACT enabled for both jobs compared to without REACT, under different degrees of overlap. Job 1 runs 8 node NCCL Tree ALLReduce and Job 2 runs Ring AllReduce. Both run 20MB message sizes on a Star topology. We run 5 random seeds for each overlap scenario.}
    \label{fig:tree-ring-star-overlap}
    \vspace{-12pt}
\end{figure}

We now evaluate how \sysname works when two jobs run simultaneously and affect each other due to node and network sharing. We simulate one job with running Tree AllReduce and another simultaneously running Ring AllReduce on a Star topology. We evaluate with different number of overlapping nodes ie the number of nodes shared by the two jobs ($x$-axis in Fig~\ref{fig:tree-ring-star-overlap}). We select the overlapped nodes randomly and try with 5 random seeds for each node overlap case. We run both jobs with the original collective pattern and compare it against the scenario where both jobs enable \sysname. Fig~\ref{fig:tree-ring-star-overlap} shows the performance improvement for both jobs under different overlapping scenarios.

\begin{figure}[t]
\centering
    \begin{subfigure}[b]{0.48\linewidth}
       \includegraphics[width=1\linewidth]{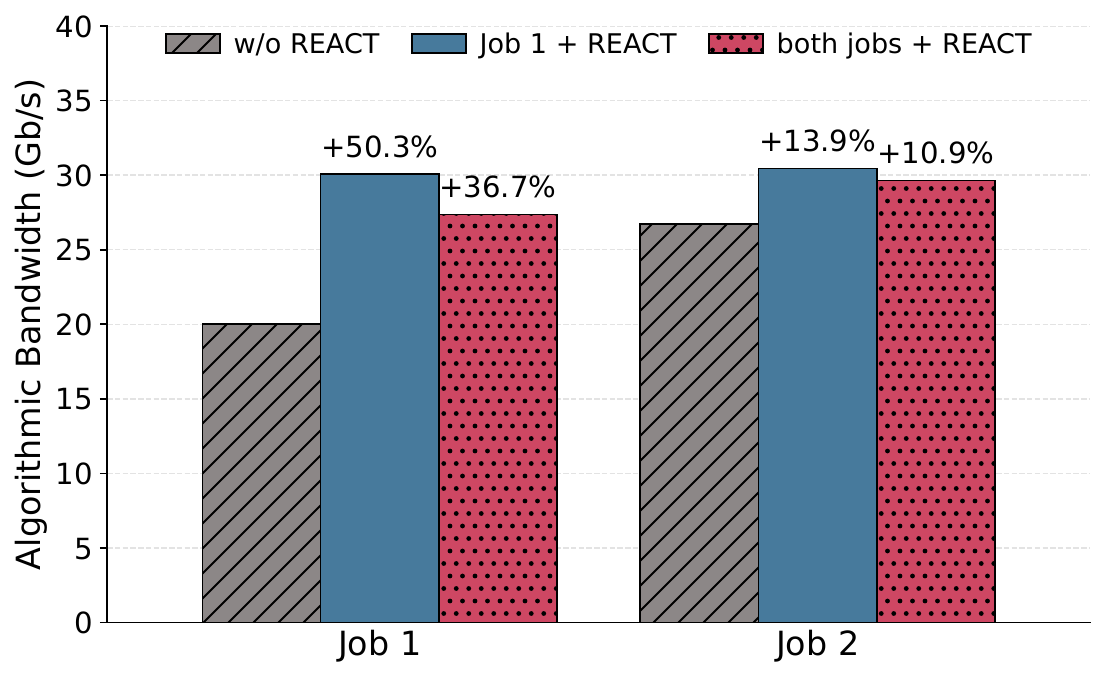}
       \vspace{-18pt}
       \caption{2 Ring AllReduce jobs}
       \label{subfig:ring-ring}
    \end{subfigure}
    ~
    \begin{subfigure}[b]{0.48\linewidth}
       \includegraphics[width=1\linewidth]{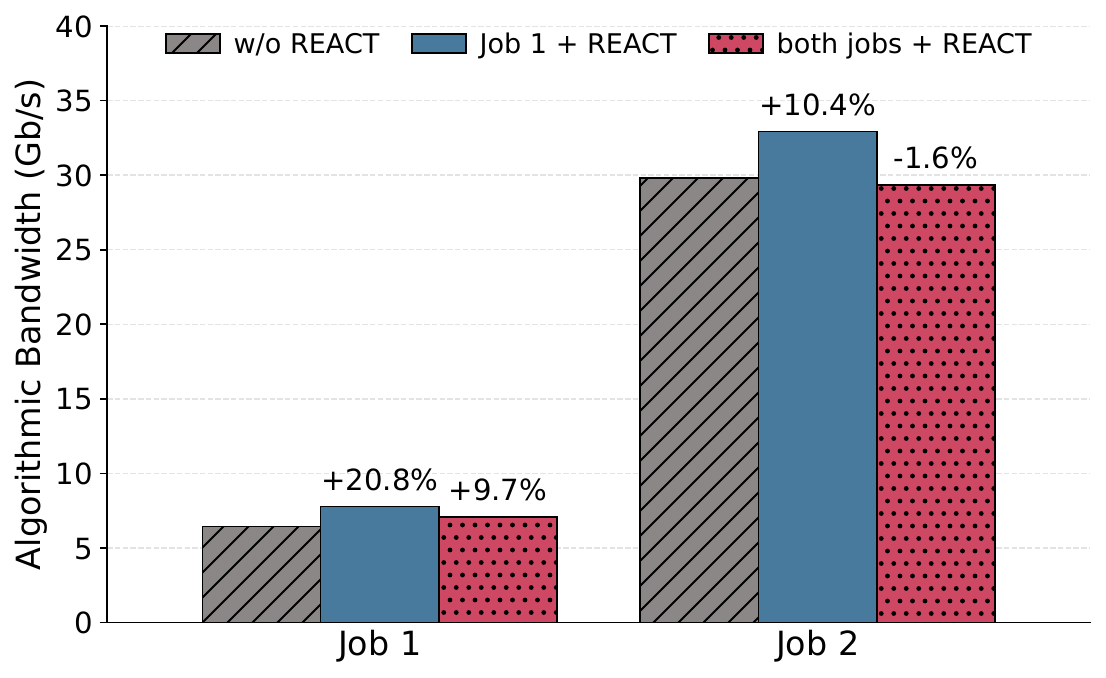}
       \vspace{-18pt}
       \caption{Recursive D. and Ring jobs}
       \label{subfig:rc-ring}
    \end{subfigure}
    
    \vspace{-11pt}
    \caption{Running 2 jobs simultaneously from Alibaba trace with only network sharing (Spine-ToR congestion). (a) runs two Ring AllReduce jobs simultaneously from Alibaba trace. (b) runs two jobs- one Recursive Doubling algorithm and other with Ring AllReduce- simultaneously. We compare performance when REACT is only enabled for Job 1 and when it is enabled for both jobs compared to without REACT (gray). The numbers on the bar are the percent change from respective baselines.}
    \label{fig:spine-barplot}
    \vspace{-12pt}
\end{figure}

Fig~\ref{fig:spine-barplot} shows two jobs simulated to run simultaneously on the network topology from Alibaba trace~\cite{alibaba-trace} with only network being shared, and no nodes overlap (Spine-ToR congestion). It shows two evaluation scenarios where the two jobs run different communication collectives, thus having different network flow profiles during runtime. 
We first enable REACT only for Job 1 and see a significant improvement in performance for both jobs -- enabling REACT can help background traffic as well.
Even in spine-ToR congestion scenarios, \sysname is able to provide significant improvement.

Next, we enable REACT for both jobs in both scenarios and we observe the performance dips compared to 1 job case -- this is due to REACT on both jobs interacting independently and getting stuck in bad configuration compared to just 1 job case. Nonetheless, we still see notable improvement in performance due to \sysname.
We leave making REACT interact constructively for all jobs as future work.


\vspace{-3pt}
\subsubsection{\sysname under Network Failures}
\label{subsubsec:network-failures}
\vspace{-3pt}

\begin{figure}[t]
\centering
       \includegraphics[width=1\linewidth]{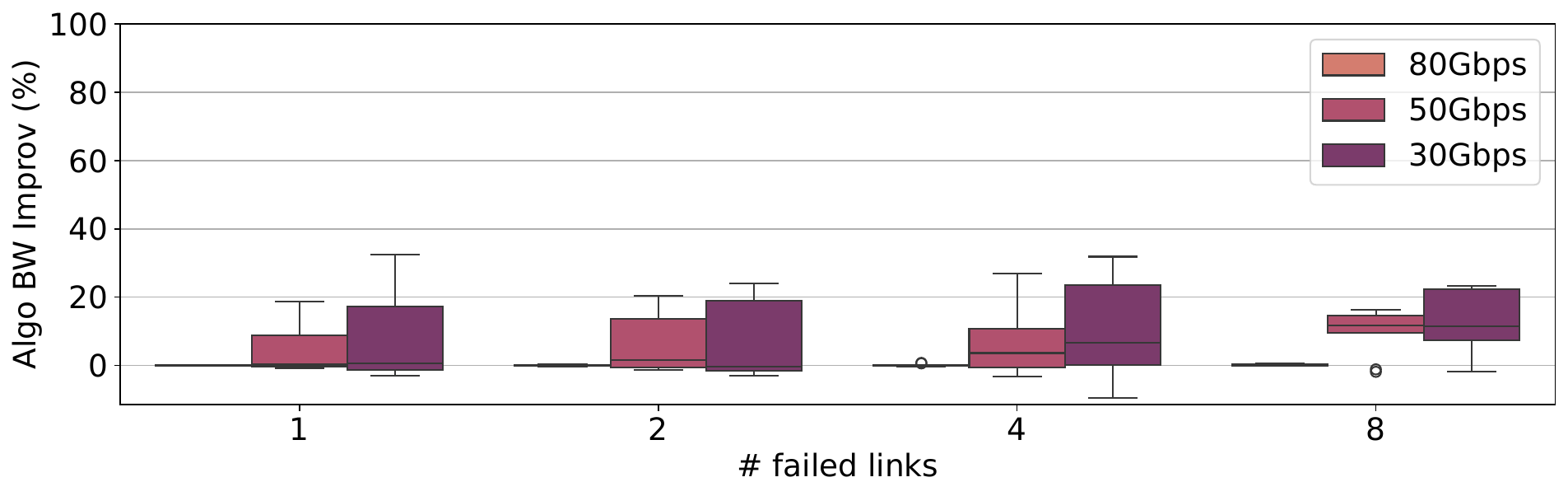}
    \vspace{-22pt}
    \caption{8 node nccl Tree Allreduce, 20MB message size, Fattree topology, partial link failures.}
    \label{fig:partial-link}
    \vspace{-17pt}
\end{figure}

We simulate gray/partial link failures~\cite{flock, reps} in the 3-layer Clos topology by reducing the link bandwidth of randomly chosen links from anywhere in the three layers to 30Gbps, 50Gbps and 80Gbps. We still use 4-way ECMP so the bandwidth of any flow over the epochs does not directly reduce to the failed link.
Fig~\ref{fig:partial-link} shows the algorithm bandwidth improvement due to \sysname for the Tree AllReduce collective-- in the worst case, it can help improve performance by upto 20\%.

%% file: discussion.tex
\vspace{-5pt}
\section{Discussion and Limitations}
\label{sec:discussion}
\vspace{-5pt}


\paragraphb{Hierarchical Collectives}
\ML training jobs may use different hierarchies of collectives to implement 3D parallelism~\cite{megatron-lm,megatron} or just divide communication into macro steps. In such cases, we can treat each level of collectives as independent jobs and apply \sysname to it, treating other collective calls within the larger job as independent smaller \ML jobs.

\paragraphb{Congestion Scenarios}
We explore unknown network paths when we swap nodes in a collective pattern, this will only work if only a limited number of nodes are congested at a time. If most links are congested, no replacement or tuning approach will work to avoid congestion.

\paragraphb{Intra-host congestion and collectives} Beyond the scenarios discussed above, \sysname can also be extended to address issues such as stragglers and host congestion. Although we do not evaluate these cases in this work, if a node is observed to be straggling, it can be moved to later timesteps of the collective to avoid delaying earlier timesteps.
\sysname can also be extended to intra-server collectives to react to observed host congestion, as well as to support other popular collective operations beyond those discussed in this work.

\paragraphb{Reducing overheads} 
Prior works already talk about reducing overheads for RDMA connection establishment (e.g.~\cite{UCMFastRDMA}). We can leverage such works along with modifying NCCL to implement \sysname with lower overheads.




%% file: conclusion.tex
\vspace{-3pt}
\section{Conclusion}
\label{sec:conclusion}
\vspace{-3pt}


In this work, we present \sysname, a system that dynamically adapts communication collective patterns at runtime to mitigate network congestion for \ML training workloads. 
It changes the sources and destinations of flows through transforms of swapping nodes in the collective graph, while ensuring that the initial data exchange operation remains semantically correct. 
\sysname operates using only end-host observations, avoiding congested links, without requiring network support or global coordination.
We evaluate \sysname on a shared cluster, demonstrating $13$-$35\%$ improvements in algorithm bandwidth, and up to $75\%$ gains in simulation across diverse congestion scenarios.
More broadly, our results highlight that restructuring application-level dataflow is a practical and complementary alternative to traditional rate control and routing for managing congestion in distributed training.


%% file: appendix.tex
\appendix
\setcounter{subfigure}{0}

\section{More Details on Collective and Transforms}

\subsection{Communication Collective patterns}
\label{sec:nccl-collectives}

We briefly explain below the collectives we target with \sysname:

\noindent\textbf{Ring AllReduce:} A \emph{Ring} based AllReduce~\cite{ring-allreduce, all-coll-algos} connects all $N$ nodes in a ring and divides the message data $M$ into $N$ equal sized chunks. At each timestep, each node sends a chunk to its left neighbor and receives one from its right neighbor. In $N-1$ steps, all nodes have reduced one chunk of data, hereby completing a reduce-scatter. In next $N-1$ steps, the ring completes an AllGather for the reduced chunks. The \emph{Ring AllGather} is also implemented in a similar way but the complete message is sent in each timestep and it only requires $N-1$ steps.

\noindent\textbf{Tree AllReduce:} A \emph{Tree} based AllReduce~\cite{nccltreeallreduce, tree-allreduce} arranges the nodes in a binary tree format, Fig~\ref{subfig:oddtree}, and data is exchanged along the edges. At the end of the initial exchange, all data is reduced at the root of the tree. Next, the data is broadcast along the reverse tree following down from the root. This algorithm is further optimized by dividing the message $M$ into two chunks and constructing two logical trees, as shown in Fig~\ref{fig:example}. This is the dual binary tree as implemented by NCCL. The benefit of using a ring is that it uses $O(N)$ steps but is bandwidth optimal as all nodes continuously send data. Tree based AllReduce has $O(\log N)$ steps and scales better~\cite{nccltreeallreduce}.

\noindent\textbf{Recursive Doubling AllGather:}
A Recursive doubling AllGather~\cite{all-coll-algos} executes $\log N$ steps, in each timestep $s$, a node with rank $r$ sends data to $q=r \texttt{ XOR } 2^s$. In each time step, the size of data sent is doubled. Generally, recursive doubling only works if $N$ is a power of 2 but it can be adapted to work for other values of $N$. 

\subsection{Applying Transforms on Collectives}
\label{sec:appendix:ring-rc-transforms}

We apply and evaluate the transforms from \S\ref{subsec:safe-transform} for Ring AllReduce and Recursive Doubling AllGather below. 
Table~\ref{tab:transforms-summary} summarizes the transforms and types of congestion each transform is able to alleviate.

\noindent\textbf{Tree AllReduce:} As discussed in \S\ref{subsec:safe-transform}, all three transforms are able to alleviate the various congestion scenarios.



\begin{figure}[t]
\centering
    \begin{subfigure}[b]{0.3\linewidth}
       \includegraphics[width=1\linewidth]{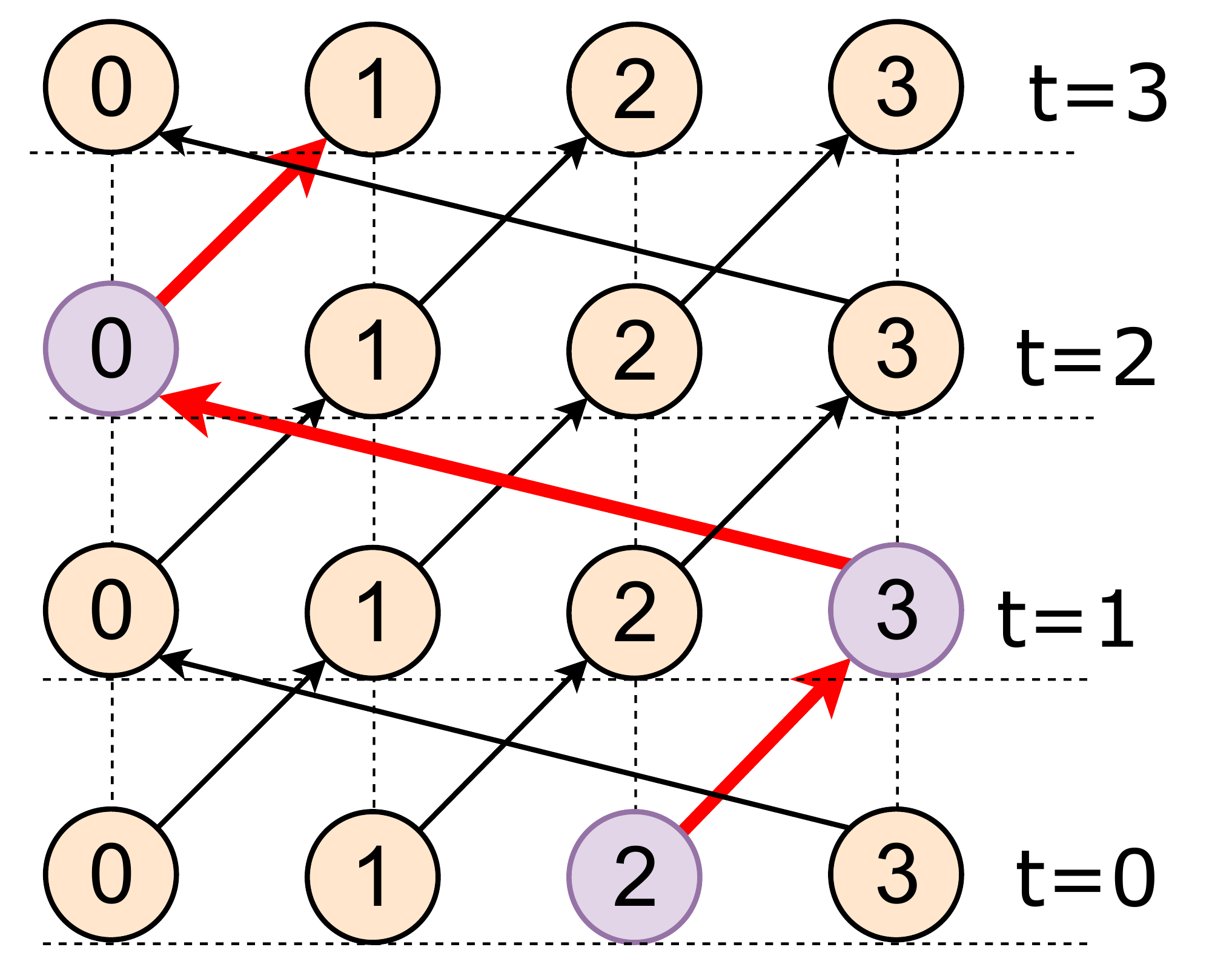}
       \caption{Ring (TEN)}
       \label{subfig:ten-ring}
    \end{subfigure}
    ~
    \begin{subfigure}[b]{0.37\linewidth}
       \includegraphics[width=1\linewidth]{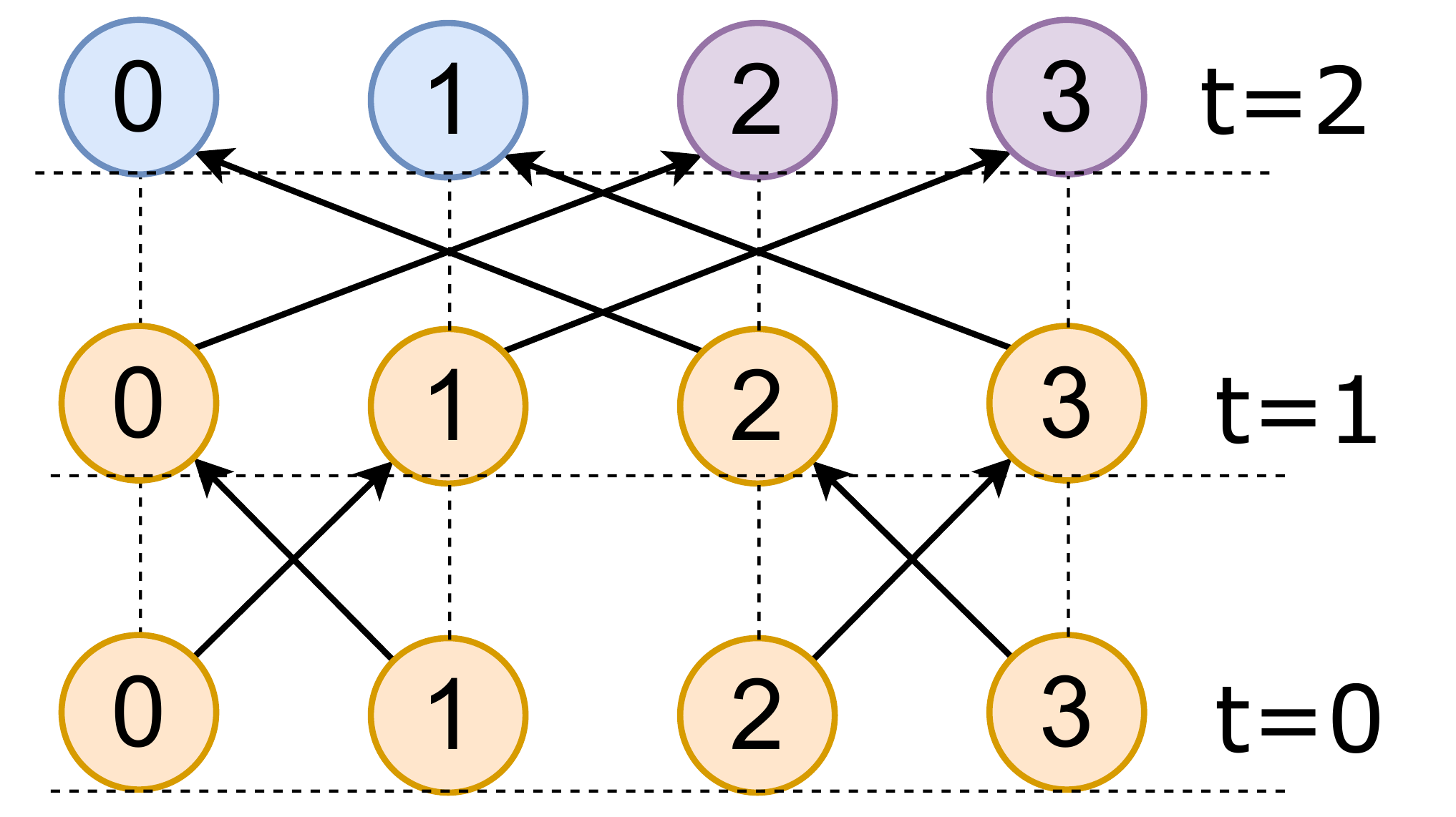}
       \caption{Recursive D.}
       \label{subfig:ten-rc}
    \end{subfigure}
    ~
    \begin{subfigure}[b]{0.2\linewidth}
       \includegraphics[width=1\linewidth]{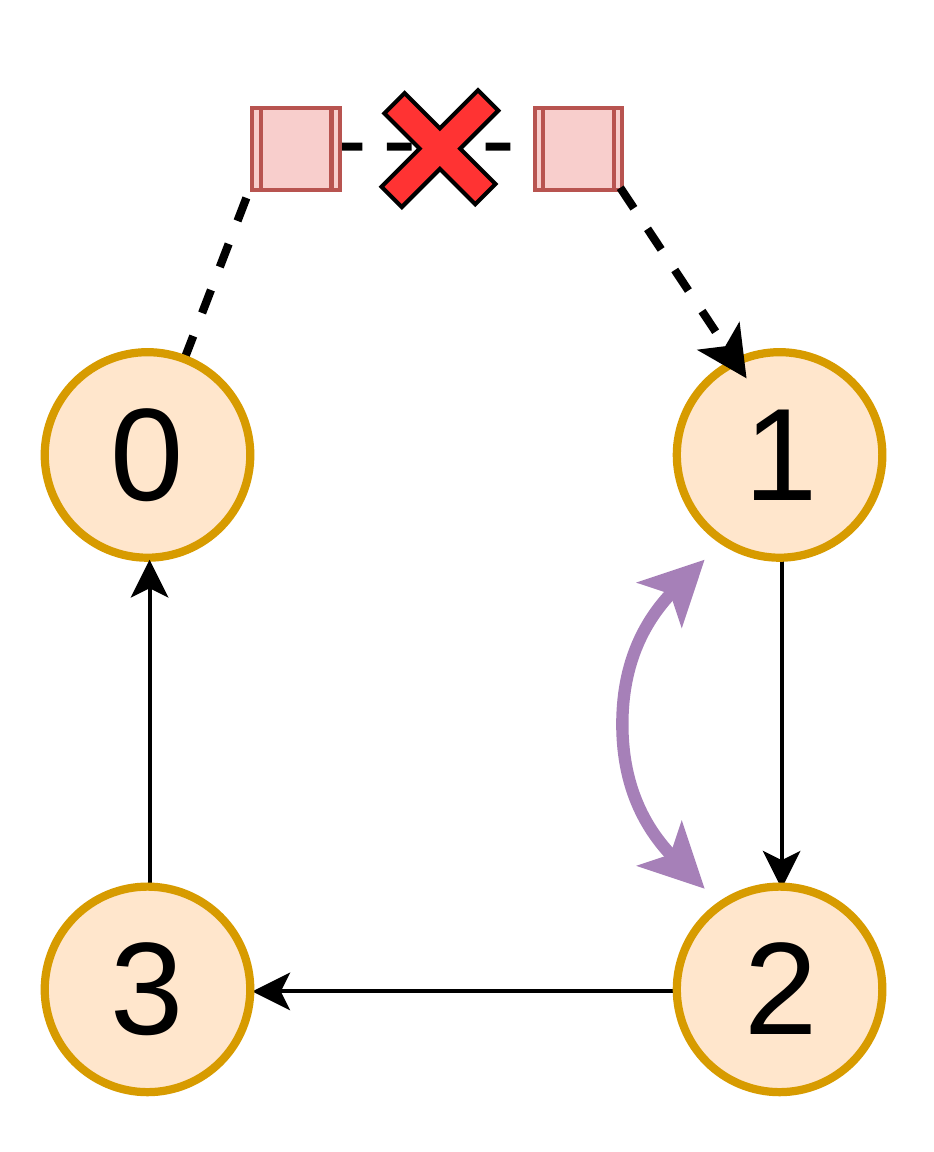}
       \caption{Ring}
       \label{subfig:ring-spine-cong} 
    \end{subfigure}
    \vspace{-12pt}
    \caption{Ring (Fig~\ref{subfig:ten-ring}) and Recursive Doubling Collective (Fig~\ref{subfig:ten-rc}) patterns. Fig~\ref{subfig:ring-spine-cong} shows a Global permutation Swap for Ring AllReduce. If $0\rightarrow 1$ is congested at the link between two red switches, swapping 1 and 2 can alleviate congestion.}
    \label{fig:ten graphs}
    \vspace{-15pt}
\end{figure}

\noindent\textbf{Ring AllReduce:} Fig~\ref{subfig:ten-ring} shows that when the $0\rightarrow 1$ link experiences Spine-ToR congestion, swapping all instances of node 1 with node 2 can help alleviate it (as per the global swap transform {~\textcircled{1}}). Since all nodes in the ring have one incoming and one outgoing flow, if $0\rightarrow 1$ flow is Server-TOR congested, then no swaps are helpful as the in and out-degrees remain the same. 

The red edges in Fig~\ref{subfig:ten-ring} highlight one of the chunk-graphs present, with violet nodes highlighting equivalent swappable nodes. Swapping nodes within the chunk graph (\textcircled{2}, \textcircled{3}) is not helpful in the Ring AllReduce as nodes in the final \collsched will end up with a higher degree (swapping 0 (violet) with 3 (violet) will lead to higher degree for nodes 0 and 3 at time steps 1 and 2 respectively).

\noindent\textbf{Recursive Doubling AllGather} incurs larger message sizes in later timesteps. Global Swap Transform {~\textcircled{1}} helps alleviate congestion by assigning lower-cost flows to later timesteps, at the expense of assigning higher-cost flows to earlier timesteps when message sizes are smaller. Since there is only one chunk graph, Chunk-Swap Transform {~\textcircled{2}} is already addressed by {~\textcircled{1}}.
The blue and purple nodes in Fig.~\ref{subfig:ten-rc} represent two equivalent swappable sets (\textcircled{3}) for \collsched, which can help mitigate spine-ToR congestion scenarios (i.e., \irregular congestion).



\subsection{Analyzing Position-equivalent Swaps}
\label{subsec:appendix:brute-force-analysis}
We briefly explain the brute force algorithm used to find all Position-equivalent swaps for a given TEN graph and its complexity.

We assign all nodes a one hot-encoded bitvector with a single bit set to 1 based on rank. We simulate an execution of the collective operation on this vector by running a breadth-first traversal of the collective graph and checking whether the final bitvector on all nodes aligns with the conditions of the collective operation. For a graph $G(V,E)$ with vertices $V$ and edges $E$, this traversal will take $O(|V|+|E|)$ time. If after the traversal, all nodes have all bits (or corresponding bits for the collective) as 1, we know the collective is semantically correct. For all pairwise vertices in the TEN, we swap them and conduct this brute force safety check. This gives a complexity of $O(|V|^2(|V|+|E|))$. For a general collective pattern with $N$ ranks and $T$ steps, $|V|=N\cdot T$, and for general collectives $T=O(N)$. Hence the complexity could reach up to $O(N^7)$. But for the popular and high performing collectives that are paretto-optimal~\cite{sccl}, we observe TEN has $|V|+|E|=O(N\log N)$. Hence, for popular collectives we have the complexity as $O((N\log N)^3)$.

\subsection{Summarizing \sysname Parameters}

Table~\ref{tab:parameters} gives a summary of all parameters used by \sysname.

\begin{table}[h!]
\small
    \centering
\vspace{-2mm}
    \begin{tabular}{|p{0.18\columnwidth} | p{0.72\columnwidth}|}
    \hline
        \textbf{Parameters} & \textbf{Definition} \\
        \hline
        $B$ & Batch size of epochs\\
        $\delta_1$ & Steady congestion parameter (\S\ref{subsec:detect-congestion}) \\
        $\delta_2$ & irregular congestion parameter (\S\ref{subsec:detect-congestion}) \\
        $K_t$ & Number of trials each turn \\
        $\epsilon$ & Threshold of min. expected perf. improvement \\
        $H$ & How much older data we maintain \\
        $E_{max}$ & Number of exploratory rounds\\
        $A_{max}$ & Maximum exploration attempts \\
        $\tau_r$ & Revert threshold to measure perf. degradation\\
        

    \hline
    \end{tabular}
    \vspace{-2mm}
    \caption{\small{Parameters used by \sysname}}
    \label{tab:parameters}
    \vspace{-5.5mm}
\end{table}

\subsection{Feedback Loop Algorithm}
\label{sec:appendix:transforms}

Algorithm~\ref{algo:swap-node} gives the algorithm we use in \S\ref{subsubsec:replace-nodes-loop} to apply the swapping transforms to collective patterns.

\begin{algorithm}
\SetAlgoLined
\DontPrintSemicolon
\KwIn{$ G:\{\langle V_c, E_c \rangle\: c\in C\} $ {TEN graph of algorithm}}
\KwIn{$ \texttt{FCT} $ {FCTs of all flows}}
\KwIn{$ \mathcal{D} $ {Old FCTs of all flows}}
\KwIn{$ \texttt{r} $ {local rank of device}}

    \SetKwFunction{FMain}{ApplyTransforms}
    \SetKwProg{Fn}{Function}{:}{}
    \Fn{\FMain{}}{
        $P \longleftarrow \FuncSty{GetCriticalPaths}(G, \texttt{FCT})$;
        
        $C \longleftarrow \FuncSty{DetectCongestion}(\texttt{FCT})$;

        $X\longleftarrow C\cap P$

        $X\longleftarrow \FuncSty{SortCongestedNodes}(X)$

        $u, c \longleftarrow 1,1$

        
        \ForEach{$ x \leftarrow \texttt{pop}(X) \text{ \& } u\texttt{++}<K_t \text{ \& } c<K_{max}$}       
        {

            $V \longleftarrow \FuncSty{ViableSet}(x, G)$

            $r \longleftarrow \FuncSty{Random}(V)$

            \While{$\texttt{True} \text{ and } u\texttt{++}<K_{t}$}
            {
                $V\longleftarrow V-\{r\}$
                
                $G'\longleftarrow \FuncSty{Swap}(G, x, r)$

                \CommentSty{// compare expected performance}
                
                \CommentSty{// of $G'$ vs $G$}
                
                \If{$E_\mathcal{D}(G')>(1+\epsilon)E_\mathcal{D}(G)$}
                {
                    $G\longleftarrow G'$

                    $X\longleftarrow \FuncSty{GetCriticalPaths}(G', \mathcal{D})\cap C$

                    $c\longleftarrow c+1$
                    
                    \texttt{break}
                }

                $r \longleftarrow \FuncSty{Random}(V)$
            }
            
        }
        \textbf{return} $ G; $ 
    }

\textbf{End Function}
\vspace{5pt}
\caption{Algorithm to execute swapping transforms.}
\label{algo:swap-node}

\end{algorithm}

\section{Implementation Details}
\label{sec:impl-additional-details}

\begin{figure}
\centering
    \begin{subfigure}[b]{0.38\linewidth}
       \includegraphics[width=1\linewidth]{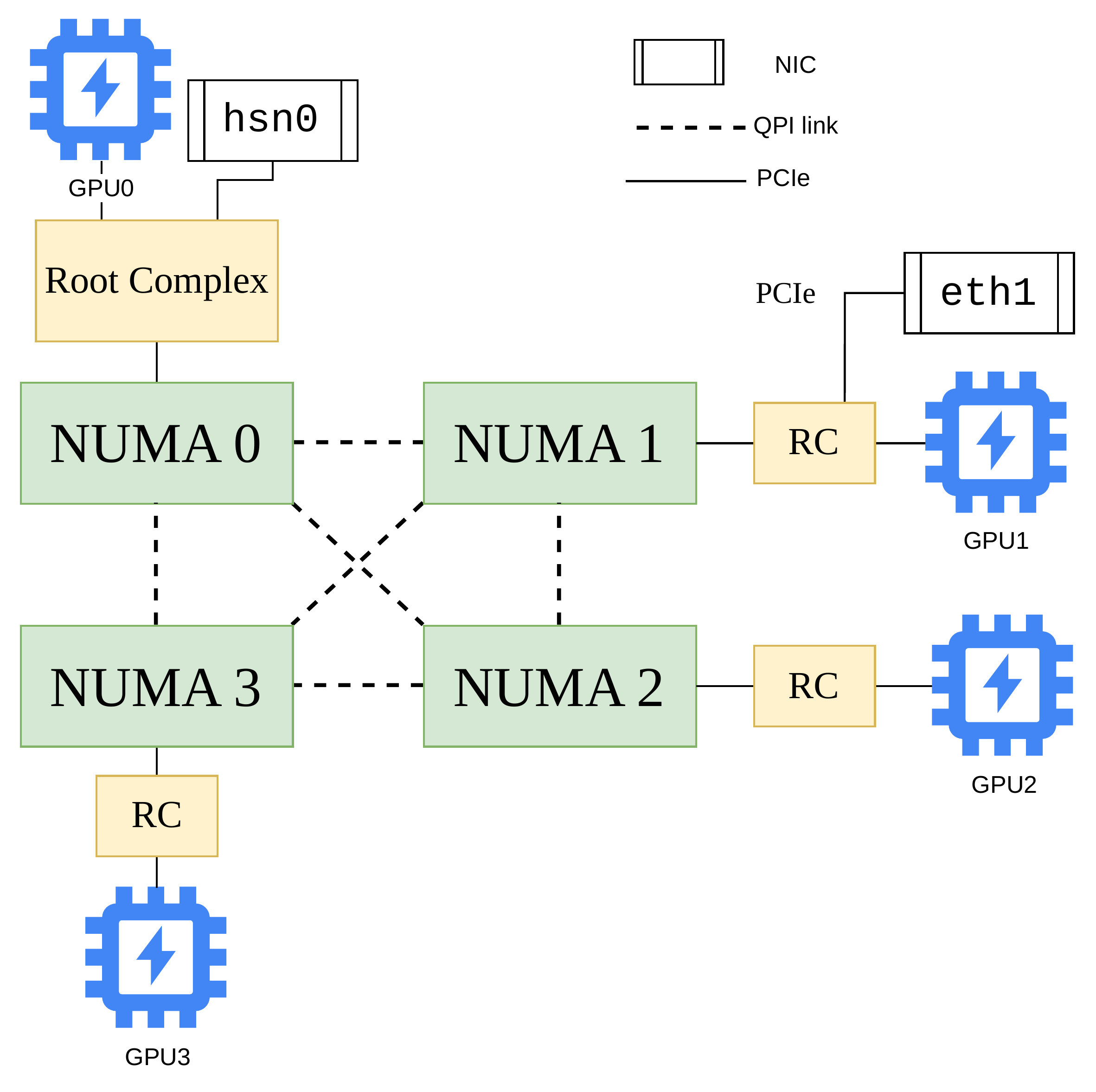}
       \caption{Intra-node topology.}
       \label{subfig:intra-node}
    \end{subfigure}
    ~
    \begin{subfigure}[b]{0.5\linewidth}
       \includegraphics[width=1\linewidth]{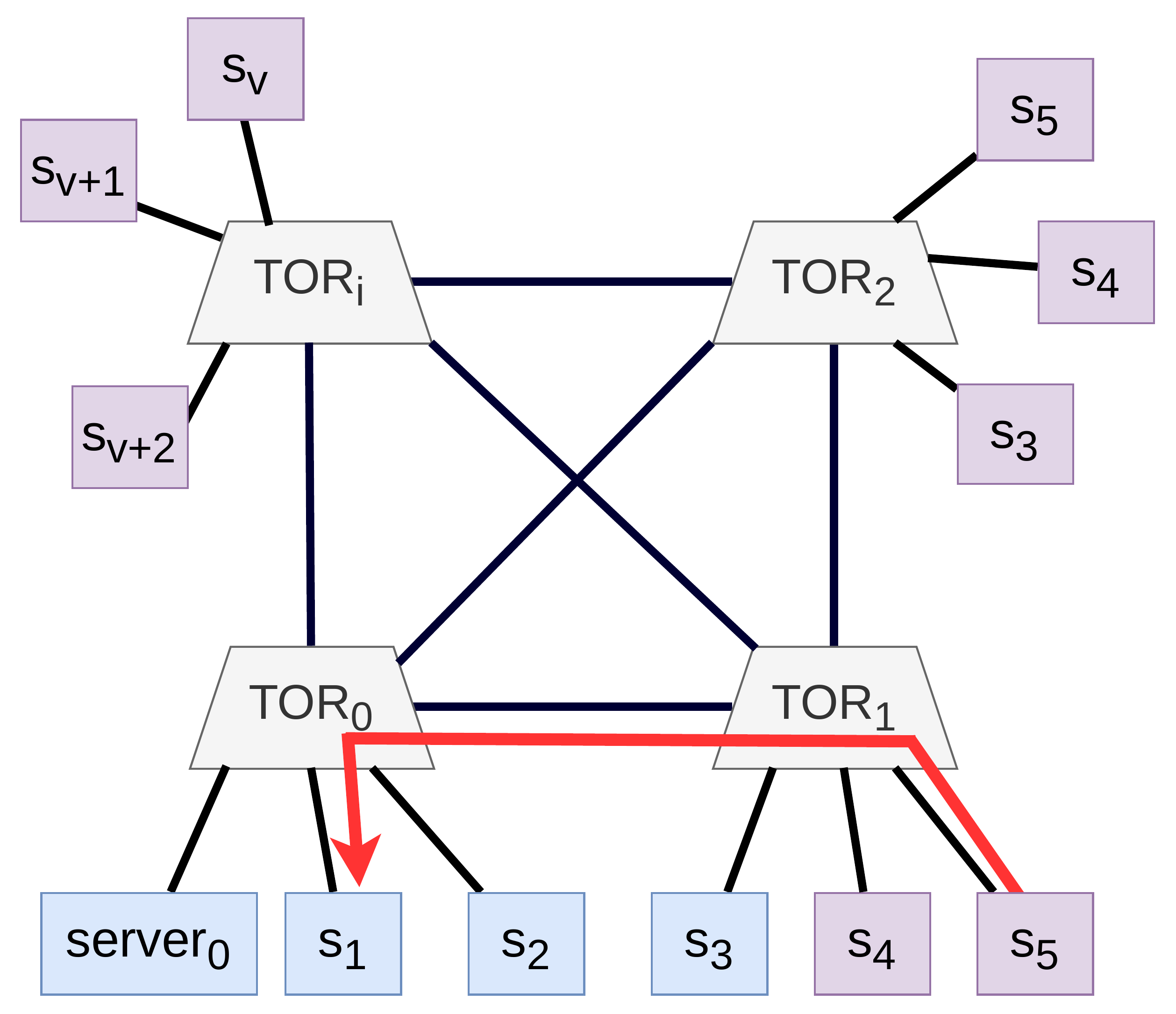}
       \caption{Slingshot network topology.}
       \label{subfig:network-topo}
    \end{subfigure}
    \caption{The intra-node and network topology in the shared academic GPU cluster. Within a server, four A100 GPUs are interconnected by NVLink (not shown for clarity). The ToRs in Slingshot network topology are all fully connected with each other. The \texttt{hsn}-high speed interface on each server is connected to exactly one TOR. Blue nodes show an example 4 node job, red arrows marks an external congested flow injected to model congestion.}
    \label{fig:topo-delta}
\end{figure}

Fig~\ref{fig:topo-delta} shows the intra-node topology and the Cray Slingshot~\cite{cray} network topology as deployed in the shared cluster. We run our experiments with GPU Direct RDMA enabled. The blue nodes in Fig~\ref{subfig:network-topo} show an example ML job running on 4 nodes and the red line is an injected congestion flow, as we inject in \S\ref{subsec:testbed}. 

\paragraphb{Adaptive Routing}
Note that we also enable adaptive routing in our deployment.
The red flow is preferably routed along the shortest path as shown. In case $\texttt{TOR}_1\rightarrow \texttt{TOR}_0$ link is congested, the switch can autonomously decide to load balance the red flow to route via $\texttt{TOR}_2$. Note that \sysname does not modify/control this routing decision but works regardless of this load balancing scheme.


\subsection{Implementation of CTCollective}
\label{subsec:ctcoll-algo}

Algorithm~\ref{algo:ctcollective} shows the pseudo code we use to implement the 
\begin{algorithm}
\SetAlgoLined
\DontPrintSemicolon
\KwIn{$ G:\{\langle V_c, E_c \rangle\: c\in C\} $ {TEN graph of algorithm}}
\KwIn{$ \texttt{r} $ {local rank of device}}

    \SetKwFunction{FMain}{RunCollective}
    \SetKwProg{Fn}{Function}{:}{}
    \Fn{\FMain{$D$}}{
        $m \longleftarrow \FuncSty{MaxDegree}(\texttt{r}, G)$;
        
        $S \longleftarrow \FuncSty{CreateStreams}(|C|\times m)$;

        $G_r \longleftarrow \FuncSty{LocalGraph}(\texttt{r}, G)$
        
        \ForEach{$ c \in C $}        
        {
            \CommentSty{// for each chunk sub-graph}
            
            \ForEach{$t \in G_r$}
            {
                \ForEach{$e \in E(c,t,\texttt{r})$}
                {
                    $\texttt{src}, \texttt{dst}\longleftarrow e$

                    \If{$\texttt{r}==\texttt{src}$}
                    {
                        \CommentSty{stream $S[c\cdot m + e]$:}
                    $  \FuncSty{post\_to\_gpu}(\texttt{isend}(\texttt{dst}))$
                    
                    }\Else{
                        \CommentSty{stream $S[c\cdot m + e]$:}
                    $  \FuncSty{post\_to\_gpu}(\texttt{irecv}(\texttt{src}))$
                    }
                }
                $\FuncSty{block\_streams}(S[c\cdot m:(c+1)\cdot m-1])$

                \If{$\texttt{agg}\in E(c,t,\texttt{r})$}
                {
                    \CommentSty{stream $S[c\cdot m]$:}$\FuncSty{post\_to\_gpu}$(\texttt{agg})
                }
                $\FuncSty{block\_streams}(S[c\cdot m:(c+1)\cdot m-1])$
            }
        }
        \textbf{return} $ F; $ 
    }

\textbf{End Function}

\caption{Function for implementing any custom graph in CTCollective library}
\label{algo:ctcollective}

\end{algorithm}

\section{Additional Evaluation Results}
\label{sec:addl-eval}

\subsection{Testbed evaluation}
\label{subsec:comp-nccl}

\begin{figure}
\centering
    \begin{subfigure}[b]{0.48\textwidth}
       \includegraphics[width=1\linewidth]{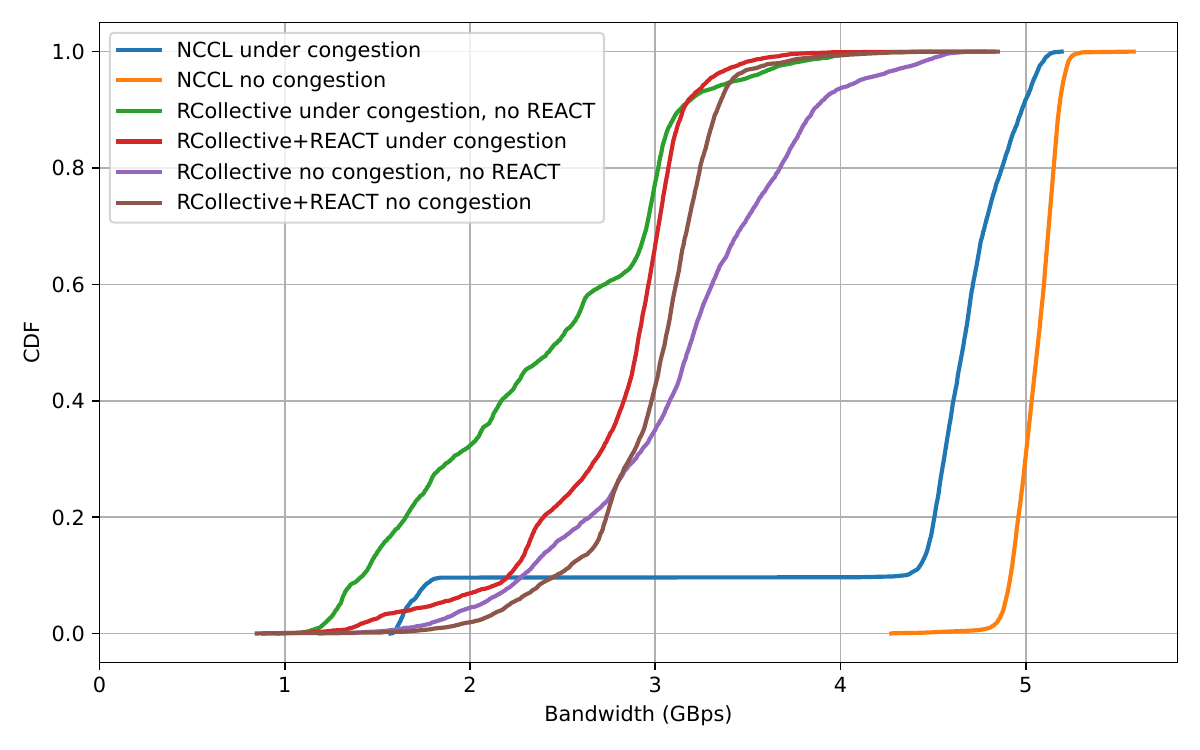}
       \caption{\sysname under $p=2$ congestion.}
       \label{subfig:rt-vs-nccl-d1}
    \end{subfigure}
    
    \begin{subfigure}[b]{0.48\textwidth}
       \includegraphics[width=1\linewidth]{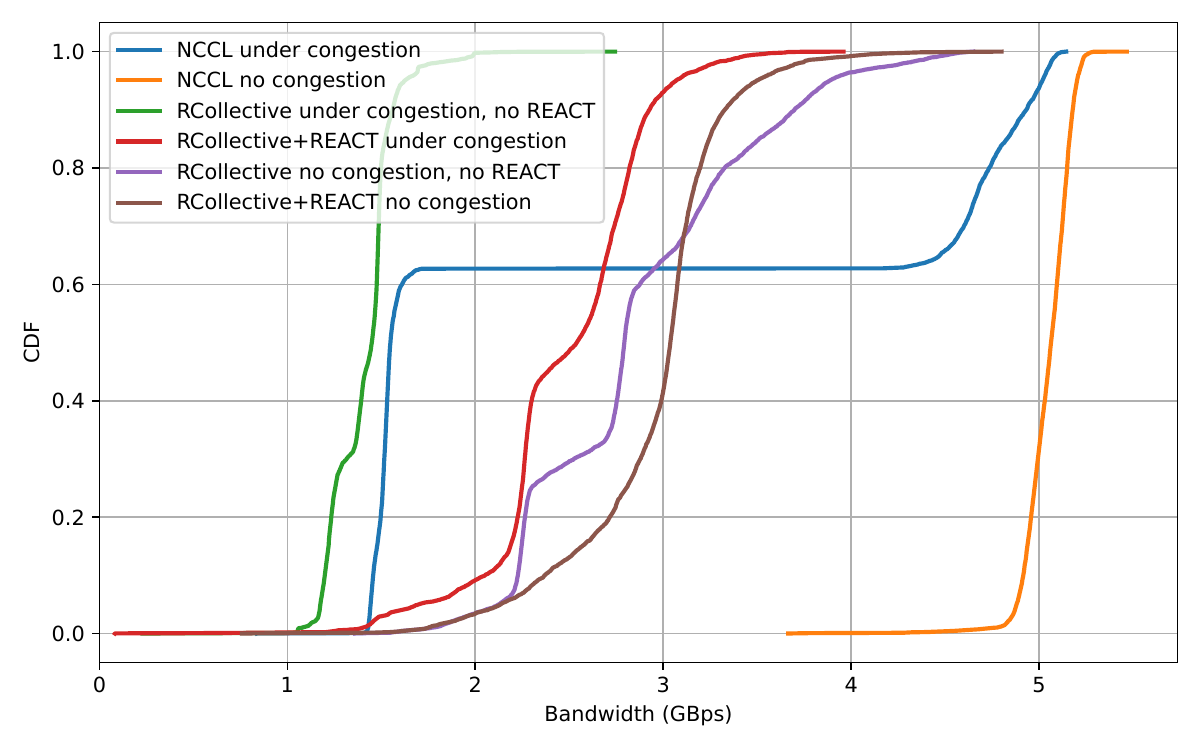}
    \caption{\sysname under $p=2$ congestion.}
    \label{subfig:rt-vs-nccl-d1}
    \end{subfigure}
    \caption{CDF of epoch times for TreeAllReduce measured on the shared GPU cluster.}
    \label{fig:nccl-vs-rt}
\end{figure}

In Fig~\ref{fig:nccl-vs-rt}, we plot the CDF of epoch times from one of the runs in \S\ref{subsec:testbed} for 4 nodes. We compare performance of baseline (RCollective with no REACT) and \sysname (RCollective+REACT) in $p=1,2$ congestion scenarios. We also plot the $p=0$ scenarios as "no congestion" lines in both plots of Fig~\ref{fig:nccl-vs-rt}. As we compare the red (\sysname) and green (baseline) lines in both plots, using \sysname, we are able to alleviate congestion. We also come very near to the purple plot (baseline under no congestion).
The brown (\sysname) line is close to the baseline (purple) under no congestion. 

\paragraphb{Comparison with NCCL}
Finally, we plot the NCCL with and without external congestion with blue and orange respectively. External congestion affects NCCL's performance in both cases. RCollective performs worse than NCCL in most cases due to the overheads of p2p communicators. Note that under $p=2$, RCollective+REACT performs better than NCCL under congestion for a majority of the epochs. 

\subsection{Results on Star Topology}

\begin{figure}[h]
\centering
       \includegraphics[width=1\linewidth]{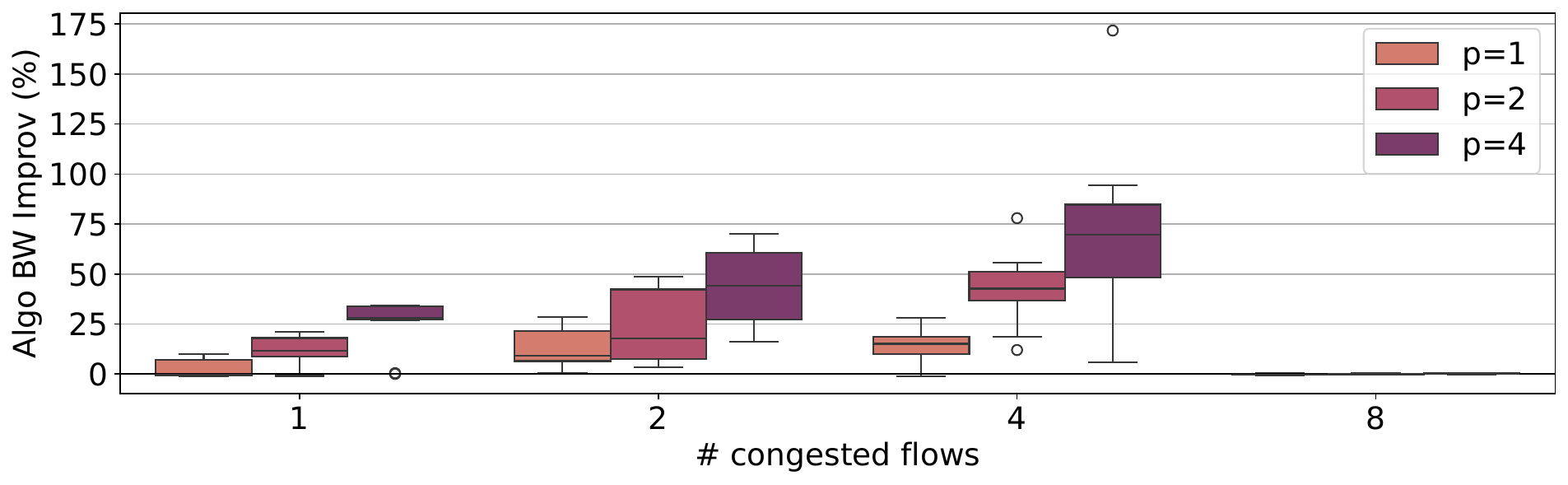}
       \caption{8 node nccl Tree Allreduce, 20MB message size, Star topology}
       \label{subfig:star-20mb}
\end{figure}

We also evaluate experiments with external congested flows for the Tree Allreduce by ns-3 simulations on the Star topology in Fig~\ref{subfig:star-20mb}. We find that higher improvements in this case than in Fattree (Fig~\ref{subfig:fattree-20mb})-- due to fixed paths in star topology, \sysname is able to make better decisions.

\subsection{Tail Latency and effects of ECMP.}
\label{subsubsec:tail-and-ecmp}

\begin{figure}
\centering
    \begin{subfigure}[b]{0.48\textwidth}
       \includegraphics[width=1\linewidth]{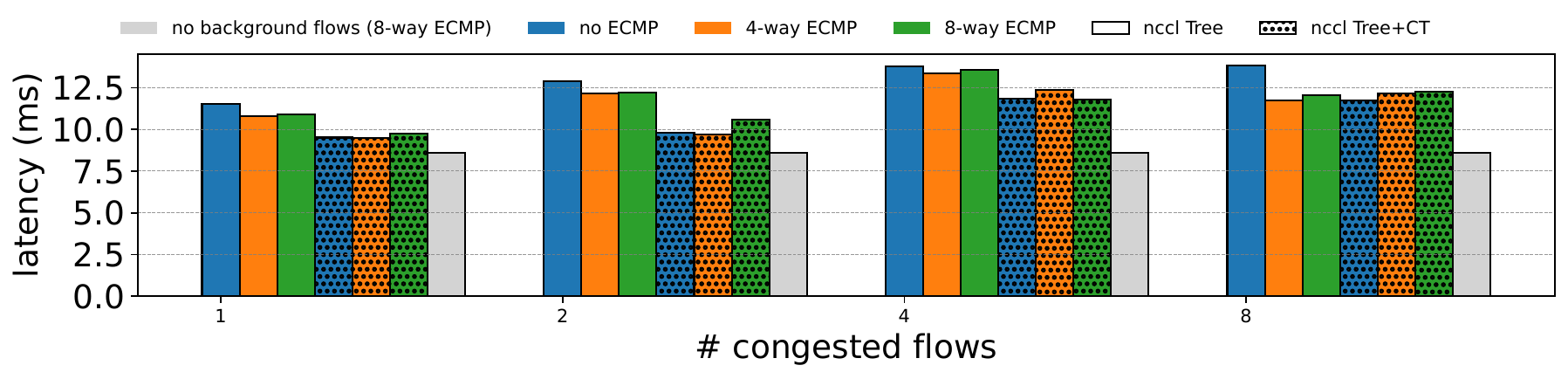}
       \caption{8 node nccl Tree Allreduce, 20MB message size, congestion degree $p=1$}
       \label{subfig:p95-fattree-20mb} 
    \end{subfigure}
    
    \begin{subfigure}[b]{0.48\textwidth}
       \includegraphics[width=1\linewidth]{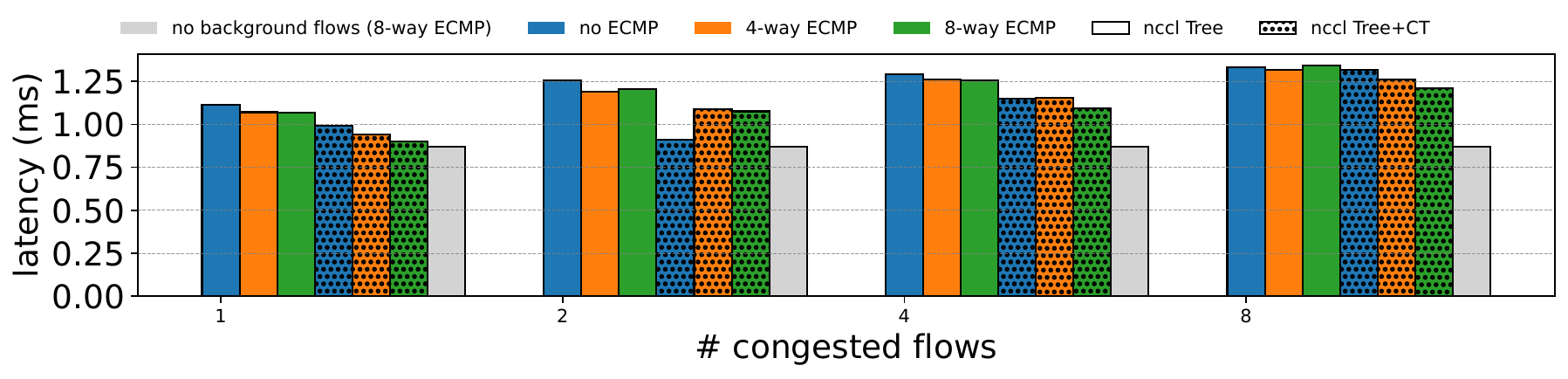}
       \caption{8 node nccl Tree Allreduce, 2MB message size, congestion degree $p=1$}
       \label{subfig:p95-fattree-2mb}
    \end{subfigure}

    \begin{subfigure}[b]{0.48\textwidth}
       \includegraphics[width=1\linewidth]{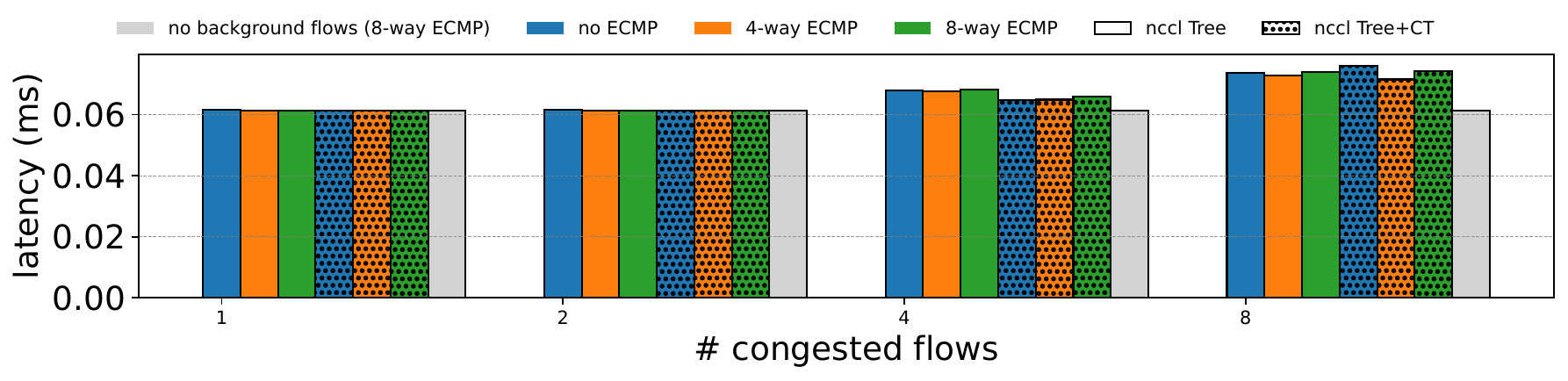}
       \caption{8 node nccl Tree Allreduce, 100KB message size, congestion degree $p=1$}
       \label{subfig:p95-fattree-100kb}
    \end{subfigure}

    \begin{subfigure}[b]{0.48\textwidth}
       \includegraphics[width=1\linewidth]{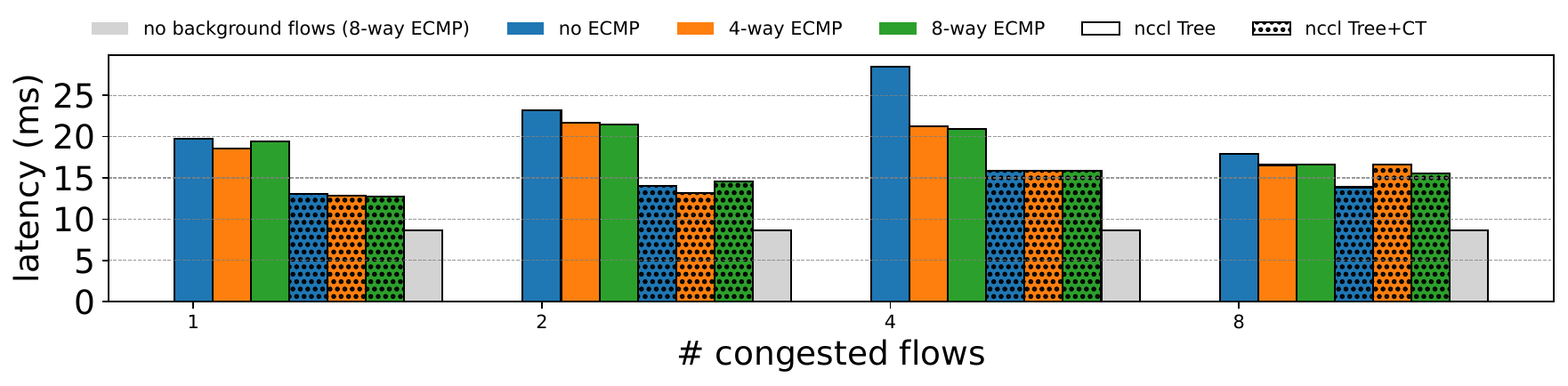}
       \caption{8 node nccl Tree Allreduce, 20MB message size, congestion degree $p=4$}
       \label{subfig:p95-fattree-20mb-cd4}
    \end{subfigure}
    
    \begin{subfigure}[b]{0.48\textwidth}
       \includegraphics[width=1\linewidth]{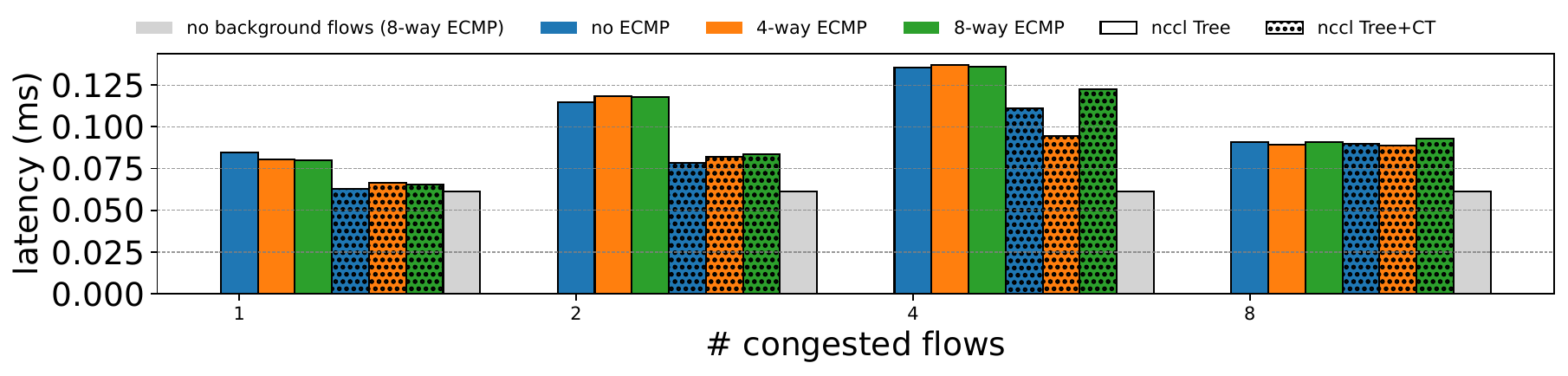}
       \caption{8 node nccl Tree Allreduce, 100KB message size, congestion degree $p=4$}
       \label{subfig:p95-fattree-100kb-cd4}
    \end{subfigure}
    \caption{Algorithm bandwidth improvement in a fattree topology for the p95 case. Comparing the effects of ECMP and collective tuning in alleviating latency deterioration due to network congestion. Increasing ECMP ports/QPs comes at additional cost while collective tuning works even with less QP configurations.}
    \label{fig:p95bw}
\end{figure}

Fig~\ref{fig:p95bw} shows the p95 epoch time across all randomly generated scenarios for number and degree of congested flows in Fig~\ref{subfig:fattree-20mb}. We run the same scenario (same network paths and random seed) under no congestion, as well as under congestion with different number of ECMP port choices, 1 and 8 respectively. We find that exploiting the multiple path choices through ECMP helps amortize any congested links in the network as we increase the number of port. Still, performance degrades of collective communication degrades and \sysname helps improve performance regardless of the underlying routing configuration/solution. 
We find that, with our short queue thresholds at the switches, collectives with smaller message sizes ($\leq 100$KB) are not significantly affected by congested flows at lower $p$ (Fig~\ref{subfig:p95-fattree-100kb}). At higher $p$, Fig~\ref{subfig:p95-fattree-100kb-cd4}, our solution still improves performance for these message sizes.

%% file: hotnets24-template.bib
@inproceedings {taccl,
author = {Aashaka Shah and Vijay Chidambaram and Meghan Cowan and Saeed Maleki and Madan Musuvathi and Todd Mytkowicz and Jacob Nelson and Olli Saarikivi and Rachee Singh},
title = {{TACCL}: Guiding Collective Algorithm Synthesis using Communication Sketches},
booktitle = {20th USENIX Symposium on Networked Systems Design and Implementation (NSDI 23)},
year = {2023},
isbn = {978-1-939133-33-5},
address = {Boston, MA},
pages = {593--612},
url = {https://www.usenix.org/conference/nsdi23/presentation/shah},
publisher = {USENIX Association},
month = apr
}

@inproceedings {mlt,
author = {Hao Wang and Han Tian and Jingrong Chen and Xinchen Wan and Jiacheng Xia and Gaoxiong Zeng and Wei Bai and Junchen Jiang and Yong Wang and Kai Chen},
title = {Towards {Domain-Specific} Network Transport for Distributed {DNN} Training},
booktitle = {21st USENIX Symposium on Networked Systems Design and Implementation (NSDI 24)},
year = {2024},
isbn = {978-1-939133-39-7},
address = {Santa Clara, CA},
pages = {1421--1443},
url = {https://www.usenix.org/conference/nsdi24/presentation/wang-hao},
publisher = {USENIX Association},
month = apr
}

@inproceedings {optireduce,
author = {Ertza Warraich and Omer Shabtai and Khalid Manaa and Shay Vargaftik and Yonatan Piasetzky and Matty Kadosh and Lalith Suresh and Muhammad Shahbaz},
title = {{OptiReduce}: Resilient and {Tail-Optimal} {AllReduce} for Distributed Deep Learning in the Cloud},
booktitle = {22nd USENIX Symposium on Networked Systems Design and Implementation (NSDI 25)},
year = {2025},
isbn = {978-1-939133-46-5},
address = {Philadelphia, PA},
pages = {685--703},
url = {https://www.usenix.org/conference/nsdi25/presentation/warraich},
publisher = {USENIX Association},
month = apr
}

@inproceedings {pollux,
author = {Aurick Qiao and Sang Keun Choe and Suhas Jayaram Subramanya and Willie Neiswanger and Qirong Ho and Hao Zhang and Gregory R. Ganger and Eric P. Xing},
title = {Pollux: Co-adaptive Cluster Scheduling for Goodput-Optimized Deep Learning},
booktitle = {15th {USENIX} Symposium on Operating Systems Design and Implementation ({OSDI} 21)},
year = {2021},
isbn = {978-1-939133-22-9},
pages = {1--18},
url = {https://www.usenix.org/conference/osdi21/presentation/qiao},
publisher = {{USENIX} Association},
month = jul
}

@misc{teccl,
      title={Rethinking Machine Learning Collective Communication as a Multi-Commodity Flow Problem}, 
      author={Behnaz Arzani and Siva Kesava Reddy Kakarla and Miguel Castro and Srikanth Kandula and Saeed Maleki and Luke Marshall},
      year={2023},
      eprint={2305.13479},
      archivePrefix={arXiv}
}

@inproceedings{dcqcn,
author = {Zhu, Yibo and Eran, Haggai and Firestone, Daniel and Guo, Chuanxiong and Lipshteyn, Marina and Liron, Yehonatan and Padhye, Jitendra and Raindel, Shachar and Yahia, Mohamad Haj and Zhang, Ming},
title = {Congestion Control for Large-Scale RDMA Deployments},
year = {2015},
isbn = {9781450335423},
publisher = {Association for Computing Machinery},
address = {New York, NY, USA},
url = {https://doi.org/10.1145/2785956.2787484},
doi = {10.1145/2785956.2787484},
booktitle = {Proceedings of the 2015 ACM Conference on Special Interest Group on Data Communication},
pages = {523–536},
numpages = {14},
location = {London, United Kingdom},
series = {SIGCOMM '15}
}

@article{tcpcongestioncontrol,
author = {Jacobson, V.},
title = {Congestion avoidance and control},
year = {1988},
issue_date = {August 1988},
publisher = {Association for Computing Machinery},
address = {New York, NY, USA},
volume = {18},
number = {4},
issn = {0146-4833},
url = {https://doi.org/10.1145/52325.52356},
doi = {10.1145/52325.52356},
journal = {SIGCOMM Comput. Commun. Rev.},
month = {aug},
pages = {314–329},
numpages = {16}
}

@inproceedings{crux,
author = {Cao, Jiamin and Guan, Yu and Qian, Kun and Gao, Jiaqi and Xiao, Wencong and Dong, Jianbo and Fu, Binzhang and Cai, Dennis and Zhai, Ennan},
title = {Crux: GPU-Efficient Communication Scheduling for Deep Learning Training},
year = {2024},
isbn = {9798400706141},
publisher = {Association for Computing Machinery},
address = {New York, NY, USA},
url = {https://doi.org/10.1145/3651890.3672239},
doi = {10.1145/3651890.3672239},
booktitle = {Proceedings of the ACM SIGCOMM 2024 Conference},
pages = {1–15},
numpages = {15},
location = {Sydney, NSW, Australia},
series = {ACM SIGCOMM '24}
}

@misc{desensi2024swingshortcuttingringshigher,
      title={Swing: Short-cutting Rings for Higher Bandwidth Allreduce}, 
      author={Daniele De Sensi and Tommaso Bonato and David Saam and Torsten Hoefler},
      year={2024},
      eprint={2401.09356},
      archivePrefix={arXiv},
      primaryClass={cs.DC},
      url={https://arxiv.org/abs/2401.09356}, 
}

@inproceedings {autoccl,
author = {Guanbin Xu and Zhihao Le and Yinhe Chen and Zhiqi Lin and Zewen Jin and Youshan Miao and Cheng Li},
title = {{AutoCCL}: Automated Collective Communication Tuning for Accelerating Distributed and Parallel {DNN} Training},
booktitle = {22nd USENIX Symposium on Networked Systems Design and Implementation (NSDI 25)},
year = {2025},
isbn = {978-1-939133-46-5},
address = {Philadelphia, PA},
pages = {667--683},
url = {https://www.usenix.org/conference/nsdi25/presentation/xu-guanbin},
publisher = {USENIX Association},
month = apr
}

@article{flock,
author = {Harsh, Vipul and Meng, Tong and Agrawal, Kapil and Godfrey, Philip Brighten},
title = {Flock: Accurate Network Fault Localization at Scale},
year = {2023},
issue_date = {June 2023},
publisher = {Association for Computing Machinery},
address = {New York, NY, USA},
volume = {1},
number = {CoNEXT1},
url = {https://doi.org/10.1145/3595289},
doi = {10.1145/3595289},
journal = {Proc. ACM Netw.},
month = jul,
articleno = {3},
numpages = {22}
}

@inproceedings{reps,
author = {Bonato, Tommaso and Kabbani, Abdul and Ghalayini, Ahmad and Papamichael, Michael and Dohadwala, Mohammad and Gianinazzi, Lukas and Khalilov, Mikhail and Achermann, Elias and De Sensi, Daniele and Hoefler, Torsten},
title = {REPS: Recycled Entropy Packet Spraying for Adaptive Load Balancing and Failure Mitigation},
year = {2026},
isbn = {9798400722127},
publisher = {Association for Computing Machinery},
address = {New York, NY, USA},
url = {https://doi.org/10.1145/3767295.3769320},
doi = {10.1145/3767295.3769320},
booktitle = {Proceedings of the 21st European Conference on Computer Systems},
pages = {225–246},
numpages = {22},
location = {McEwan Hall/The University of Edinburgh, Edinburgh, Scotland UK},
series = {EUROSYS '26}
}

@inproceedings{dctcp,
author = {Alizadeh, Mohammad and Greenberg, Albert and Maltz, David A. and Padhye, Jitendra and Patel, Parveen and Prabhakar, Balaji and Sengupta, Sudipta and Sridharan, Murari},
title = {Data center TCP (DCTCP)},
year = {2010},
isbn = {9781450302012},
publisher = {Association for Computing Machinery},
address = {New York, NY, USA},
url = {https://doi.org/10.1145/1851182.1851192},
doi = {10.1145/1851182.1851192},
booktitle = {Proceedings of the ACM SIGCOMM 2010 Conference},
pages = {63–74},
numpages = {12},
location = {New Delhi, India},
series = {SIGCOMM '10}
}

@misc{cray,
    title = {Cray Slingshot network},
    author = {HPE},
    url={https://www.hpe.com/psnow/doc/a50002546enw.html} 
}

@misc{nccl,
    title = {NVIDIA Collective Communications Library (NCCL)},
    author = {NVIDIA Corporation},
    url={https://developer.nvidia.com/nccl} 
}

@inproceedings{megatron-lm,
author = {Narayanan, Deepak and Shoeybi, Mohammad and Casper, Jared and LeGresley, Patrick and Patwary, Mostofa and Korthikanti, Vijay and Vainbrand, Dmitri and Kashinkunti, Prethvi and Bernauer, Julie and Catanzaro, Bryan and Phanishayee, Amar and Zaharia, Matei},
title = {Efficient large-scale language model training on GPU clusters using megatron-LM},
year = {2021},
isbn = {9781450384421},
publisher = {Association for Computing Machinery},
address = {New York, NY, USA},
url = {https://doi.org/10.1145/3458817.3476209},
doi = {10.1145/3458817.3476209},
booktitle = {Proceedings of the International Conference for High Performance Computing, Networking, Storage and Analysis},
articleno = {58},
numpages = {15},
location = {St. Louis, Missouri},
series = {SC '21}
}

@misc{alibaba-trace,
    title = {Alibaba GPU Cluster Dataset 2023},
    author = {Alibaba Group},
    url={https://github.com/alibaba/alibaba-lingjun-dataset-2023} 
}

@misc{gloo,
    title = {Gloo: Collective communications library},
    author = {Facebook},
    year = {2017},
    url={https://github.com/facebookincubator/gloo} 
}

@article{all-coll-algos,
author = {Rajeev Thakur and Rolf Rabenseifner and William Gropp},
title ={Optimization of Collective Communication Operations in MPICH},
journal = {The International Journal of High Performance Computing Applications},
volume = {19},
number = {1},
pages = {49-66},
year = {2005},
doi = {10.1177/1094342005051521},
URL = { 
    
        https://doi.org/10.1177/1094342005051521
    
    

},
eprint = { 
        https://doi.org/10.1177/1094342005051521
}
}

@article{ring-allreduce,
title = {Bandwidth optimal all-reduce algorithms for clusters of workstations},
journal = {Journal of Parallel and Distributed Computing},
volume = {69},
number = {2},
pages = {117-124},
year = {2009},
issn = {0743-7315},
doi = {https://doi.org/10.1016/j.jpdc.2008.09.002},
url = {https://www.sciencedirect.com/science/article/pii/S0743731508001767},
author = {Pitch Patarasuk and Xin Yuan}
}

@INPROCEEDINGS{adapcc,
  author={Zhao, Xiaoyang and Zhang, Zhe and Wu, Chuan},
  booktitle={2024 IEEE 44th International Conference on Distributed Computing Systems (ICDCS)}, 
  title={AdapCC: Making Collective Communication in Distributed Machine Learning Adaptive}, 
  year={2024},
  volume={},
  number={},
  pages={25-35},
  doi={10.1109/ICDCS60910.2024.00012}}

@misc{mpi,
    title = {Open MPI: Open Source High Performance Computing},
    year = {1999},
    url={https://www.open-mpi.org/} 
}

@misc{adaptiverouting,
    title = {Adaptive Routing},
    author = {NVIDIA Corporation},
    year = {2024},
    url={https://docs.nvidia.com/networking/display/ibclusterbringupprocedure/adaptive+routing} 
}

@misc{tensorflow,
    title = {Tensorflow: An end-to-end platform for machine learning},
    author = {Google},
    year = {2015},
    url={https://www.tensorflow.org/} 
}

@misc{pytorch,
    title = {PyTorch: Tensors and Dynamic neural networks in Python with strong GPU acceleration},
    author = {Facebook},
    year = {2016},
    url={https://pytorch.org/} 
}

@inproceedings{kfattree,
author = {Al-Fares, Mohammad and Loukissas, Alexander and Vahdat, Amin},
title = {A scalable, commodity data center network architecture},
year = {2008},
isbn = {9781605581750},
publisher = {Association for Computing Machinery},
address = {New York, NY, USA},
url = {https://doi.org/10.1145/1402958.1402967},
doi = {10.1145/1402958.1402967},
booktitle = {Proceedings of the ACM SIGCOMM 2008 Conference on Data Communication},
pages = {63–74},
numpages = {12},
location = {Seattle, WA, USA},
series = {SIGCOMM '08}
}

@article{HOCKNEY1994389,
title = {The communication challenge for MPP: Intel Paragon and Meiko CS-2},
journal = {Parallel Computing},
volume = {20},
number = {3},
pages = {389-398},
year = {1994},
issn = {0167-8191},
doi = {https://doi.org/10.1016/S0167-8191(06)80021-9},
url = {https://www.sciencedirect.com/science/article/pii/S0167819106800219},
author = {Roger W. Hockney}
}

@misc{adaptiveroutingwhitepaper,
    title = {Networking for the Era of AI: The Network Defines the Data Center},
    author = {NVIDIA Corporation},
    year = {2024},
    url={https://resources.nvidia.com/en-us-networking-ai/networking-overall?ncid=so-face-354533-vt45} 
}

@misc{nccltreeallreduce,
    title = {Massively Scale Your Deep Learning Training with NCCL 2.4},
    author = {Sylvain Jeaugey},
    year = {2019},
    url={https://developer.nvidia.com/blog/massively-scale-deep-learning-training-nccl-2-4/} 
}

@inproceedings{sccl,
author = {Cai, Zixian and Liu, Zhengyang and Maleki, Saeed and Musuvathi, Madanlal and Mytkowicz, Todd and Nelson, Jacob and Saarikivi, Olli},
title = {Synthesizing optimal collective algorithms},
year = {2021},
isbn = {9781450382946},
publisher = {Association for Computing Machinery},
address = {New York, NY, USA},
url = {https://doi.org/10.1145/3437801.3441620},
doi = {10.1145/3437801.3441620},
booktitle = {Proceedings of the 26th ACM SIGPLAN Symposium on Principles and Practice of Parallel Programming},
pages = {62–75},
numpages = {14},
location = {Virtual Event, Republic of Korea},
series = {PPoPP '21}
}

@inproceedings {syndicate,
author = {Kshiteej Mahajan and Ching-Hsiang Chu and Srinivas Sridharan and Aditya Akella},
title = {Better Together: Jointly Optimizing {ML} Collective Scheduling and Execution Planning using {SYNDICATE}},
booktitle = {20th USENIX Symposium on Networked Systems Design and Implementation (NSDI 23)},
year = {2023},
isbn = {978-1-939133-33-5},
address = {Boston, MA},
pages = {809--824},
url = {https://www.usenix.org/conference/nsdi23/presentation/mahajan},
publisher = {USENIX Association},
month = apr
}

@inproceedings{plink,
 author = {Luo, Liang and West, Peter and Nelson, Jacob and Krishnamurthy, Arvind and Ceze, Luis},
 booktitle = {Proceedings of Machine Learning and Systems},
 editor = {I. Dhillon and D. Papailiopoulos and V. Sze},
 pages = {82--97},
 title = {PLink: Discovering and Exploiting Locality for Accelerated Distributed Training on the public Cloud},
 url = {https://proceedings.mlsys.org/paper_files/paper/2020/file/eca986d585a03890a412587a2f5ccb43-Paper.pdf},
 volume = {2},
 year = {2020}
}

@inproceedings {cassini,
author = {Sudarsanan Rajasekaran and Manya Ghobadi and Aditya Akella},
title = {{CASSINI}: {Network-Aware} Job Scheduling in Machine Learning Clusters},
booktitle = {21st USENIX Symposium on Networked Systems Design and Implementation (NSDI 24)},
year = {2024},
isbn = {978-1-939133-39-7},
address = {Santa Clara, CA},
pages = {1403--1420},
url = {https://www.usenix.org/conference/nsdi24/presentation/rajasekaran},
publisher = {USENIX Association},
month = apr
}

@inproceedings{homa,
author = {Montazeri, Behnam and Li, Yilong and Alizadeh, Mohammad and Ousterhout, John},
title = {Homa: a receiver-driven low-latency transport protocol using network priorities},
year = {2018},
isbn = {9781450355674},
publisher = {Association for Computing Machinery},
address = {New York, NY, USA},
url = {https://doi.org/10.1145/3230543.3230564},
doi = {10.1145/3230543.3230564},
booktitle = {Proceedings of the 2018 Conference of the ACM Special Interest Group on Data Communication},
pages = {221–235},
numpages = {15},
location = {Budapest, Hungary},
series = {SIGCOMM '18}
}

@article{tree-allreduce,
author = {Sanders, Peter and Speck, Jochen and Tr\"{a}ff, Jesper Larsson},
title = {Two-tree algorithms for full bandwidth broadcast, reduction and scan},
year = {2009},
issue_date = {December, 2009},
publisher = {Elsevier Science Publishers B. V.},
address = {NLD},
volume = {35},
number = {12},
issn = {0167-8191},
url = {https://doi.org/10.1016/j.parco.2009.09.001},
doi = {10.1016/j.parco.2009.09.001},
journal = {Parallel Comput.},
month = {dec},
pages = {581–594},
numpages = {14}
}

@inproceedings{NIPS2014_186b3d04,
 author = {Lee, Seunghak and Kim, Jin and Zheng, Xun and Ho, Qirong and Gibson, Garth A. and Xing, Eric P.},
 booktitle = {Advances in Neural Information Processing Systems},
 editor = {Z. Ghahramani and M. Welling and C. Cortes and N. Lawrence and K. Weinberger},
 pages = {},
 publisher = {Curran Associates, Inc.},
 title = {On Model Parallelization and Scheduling Strategies for Distributed Machine Learning},
 url = {https://proceedings.neurips.cc/paper_files/paper/2014/file/186b3d044a8c9898679d98dbd0d9b860-Paper.pdf},
 volume = {27},
 year = {2014}
}

@inproceedings{swift,
author = {Kumar, Gautam and Dukkipati, Nandita and Jang, Keon and Wassel, Hassan M. G. and Wu, Xian and Montazeri, Behnam and Wang, Yaogong and Springborn, Kevin and Alfeld, Christopher and Ryan, Michael and Wetherall, David and Vahdat, Amin},
title = {Swift: Delay is Simple and Effective for Congestion Control in the Datacenter},
year = {2020},
isbn = {9781450379557},
publisher = {Association for Computing Machinery},
address = {New York, NY, USA},
url = {https://doi.org/10.1145/3387514.3406591},
doi = {10.1145/3387514.3406591},
booktitle = {Proceedings of the Annual Conference of the ACM Special Interest Group on Data Communication on the Applications, Technologies, Architectures, and Protocols for Computer Communication},
pages = {514–528},
numpages = {15},
location = {Virtual Event, USA},
series = {SIGCOMM '20}
}

@inproceedings{falcon,
author = {Singhvi, Arjun and Dukkipati, Nandita and Chandra, Prashant and Wassel, Hassan M. G. and Sharma, Naveen Kr. and Rebello, Anthony and Schuh, Henry and Kumar, Praveen and Montazeri, Behnam and Bansod, Neelesh and Thomas, Sarin and Cho, Inho and Seibert, Hyojeong Lee and Wu, Baijun and Yang, Rui and Li, Yuliang and Huang, Kai and Yin, Qianwen and Agarwal, Abhishek and Vaduvatha, Srinivas and Wang, Weihuang and Moshref, Masoud and Ji, Tao and Wetherall, David and Vahdat, Amin},
title = {Falcon: A Reliable, Low Latency Hardware Transport},
year = {2025},
isbn = {9798400715242},
publisher = {Association for Computing Machinery},
address = {New York, NY, USA},
url = {https://doi.org/10.1145/3718958.3754353},
doi = {10.1145/3718958.3754353},
booktitle = {Proceedings of the ACM SIGCOMM 2025 Conference},
pages = {248–263},
numpages = {16},
location = {S{\~a}o Francisco Convent, Coimbra, Portugal},
series = {SIGCOMM '25}
}

@misc{shoeybi2020megatronlmtrainingmultibillionparameter,
      title={Megatron-LM: Training Multi-Billion Parameter Language Models Using Model Parallelism}, 
      author={Mohammad Shoeybi and Mostofa Patwary and Raul Puri and Patrick LeGresley and Jared Casper and Bryan Catanzaro},
      year={2020},
      eprint={1909.08053},
      archivePrefix={arXiv},
      primaryClass={cs.CL},
      url={https://arxiv.org/abs/1909.08053}, 
}

@inproceedings{megatron,
author = {Narayanan, Deepak and Shoeybi, Mohammad and Casper, Jared and LeGresley, Patrick and Patwary, Mostofa and Korthikanti, Vijay and Vainbrand, Dmitri and Kashinkunti, Prethvi and Bernauer, Julie and Catanzaro, Bryan and Phanishayee, Amar and Zaharia, Matei},
title = {Efficient large-scale language model training on GPU clusters using megatron-LM},
year = {2021},
isbn = {9781450384421},
publisher = {Association for Computing Machinery},
address = {New York, NY, USA},
url = {https://doi.org/10.1145/3458817.3476209},
doi = {10.1145/3458817.3476209},
booktitle = {Proceedings of the International Conference for High Performance Computing, Networking, Storage and Analysis},
articleno = {58},
numpages = {15},
location = {St. Louis, Missouri},
series = {SC '21}
}

@misc{deepspeed,
      title={ZeRO: Memory Optimizations Toward Training Trillion Parameter Models}, 
      author={Samyam Rajbhandari and Jeff Rasley and Olatunji Ruwase and Yuxiong He},
      year={2020},
      eprint={1910.02054},
      archivePrefix={arXiv},
      primaryClass={cs.LG},
      url={https://arxiv.org/abs/1910.02054}, 
}

@misc{srv6-rfc8986,
    series =    {Request for Comments},
    number =    8986,
    howpublished =  {RFC 8986},
    publisher = {RFC Editor},
    doi =       {10.17487/RFC8986},
    url =       {https://www.rfc-editor.org/info/rfc8986},
    author =    {Clarence Filsfils and Pablo Camarillo and John Leddy and Daniel Voyer and Satoru Matsushima and Zhenbin Li},
    title =     {{Segment Routing over IPv6 (SRv6) Network Programming}},
    pagetotal = 40,
    year =      2021,
    month =     feb,
}

@inproceedings{ByteScheduler,
author = {Peng, Yanghua and Zhu, Yibo and Chen, Yangrui and Bao, Yixin and Yi, Bairen and Lan, Chang and Wu, Chuan and Guo, Chuanxiong},
title = {A generic communication scheduler for distributed DNN training acceleration},
year = {2019},
isbn = {9781450368735},
publisher = {Association for Computing Machinery},
address = {New York, NY, USA},
url = {https://doi.org/10.1145/3341301.3359642},
doi = {10.1145/3341301.3359642},
booktitle = {Proceedings of the 27th ACM Symposium on Operating Systems Principles},
pages = {16–29},
numpages = {14},
location = {Huntsville, Ontario, Canada},
series = {SOSP '19}
}

@inproceedings {poseidon,
author = {Hao Zhang and Zeyu Zheng and Shizhen Xu and Wei Dai and Qirong Ho and Xiaodan Liang and Zhiting Hu and Jinliang Wei and Pengtao Xie and Eric P. Xing},
title = {Poseidon: An Efficient Communication Architecture for Distributed Deep Learning on {GPU} Clusters},
booktitle = {2017 USENIX Annual Technical Conference (USENIX ATC 17)},
year = {2017},
isbn = {978-1-931971-38-6},
address = {Santa Clara, CA},
pages = {181--193},
url = {https://www.usenix.org/conference/atc17/technical-sessions/presentation/zhang},
publisher = {USENIX Association},
month = jul
}

@misc{khashab2026highspeednetworkinggigascaleai,
      title={High-speed Networking for Giga-Scale AI Factories}, 
      author={Sajy Khashab and Albert Gran Alcoz and Alon Gal and Jacky Romano and Rani Abboud and Yonatan Piasetzky and Lior Maman and Amit Nishry and Barak Gafni and Omer Shabtai and Matty Kadosh and Dror Goldenberg and Gilad Shainer and Mark Silberstein},
      year={2026},
      eprint={2605.21187},
      archivePrefix={arXiv},
      primaryClass={cs.NI},
      url={https://arxiv.org/abs/2605.21187}, 
}

@misc{mrc,
      title={Resilient AI Supercomputer Networking using MRC and SRv6}, 
      author={Joao Araujo and Alex Chow and Mark Handley and Ryder Lewis and Christoph Paasch and Jitendra Padhye and Michael Papamichael and Greg Steinbrecher and Amin Tootoonchian and Lihua Yuan and S. Anantharamu and Abhishek Dosi and Mohit Garg and Mahdieh Ghazi and Torsten Hoefler and Deepal Jayasinghe and Jithin Jose and Abdul Kabbani and Guohan Lu and Yang Wang and K. Doddapaneni and Murali Garimella and Vipin Jain and Yanfang Le and H. Nagulapalli and S. Narayanan and Rong Pan and Rathina Sabesan and Raghava Sivaramu and Rip Sohan and Eric Davis and Dragos Dumitrescu and Mohan Kalkunte and Bhaswar Mitra and Guglielmo Morandin and Adrian Popa and Costin Raiciu and Eric Spada and John Spillane and Niranjan Vaidya and Aviv Barnea and Idan Burstein and Elazar Cohen and Yamin Friedman and Noam Katz and Masoud Moshref and Yuval Shpigelman and Shahaf Shuler and Shy Shyman and Sayantan Sur},
      year={2026},
      eprint={2605.04333},
      archivePrefix={arXiv},
      primaryClass={cs.NI},
      url={https://arxiv.org/abs/2605.04333}, 
}

@inproceedings{dcp-ho-trimming,
author = {Li, Wenxue and Liu, Xiangzhou and Zhang, Yunxuan and Wang, Zihao and Gu, Wei and Qian, Tao and Zeng, Gaoxiong and Ren, Shoushou and Huang, Xinyang and Ren, Zhenghang and Liu, Bowen and Zhang, Junxue and Chen, Kai and Liu, Bingyang},
title = {Revisiting RDMA Reliability for Lossy Fabrics},
year = {2025},
isbn = {9798400715242},
publisher = {Association for Computing Machinery},
address = {New York, NY, USA},
url = {https://doi.org/10.1145/3718958.3750480},
doi = {10.1145/3718958.3750480},
booktitle = {Proceedings of the ACM SIGCOMM 2025 Conference},
pages = {85–98},
numpages = {14},
location = {S{\~a}o Francisco Convent, Coimbra, Portugal},
series = {SIGCOMM '25}
}

@inproceedings{pipedream,
author = {Narayanan, Deepak and Harlap, Aaron and Phanishayee, Amar and Seshadri, Vivek and Devanur, Nikhil R. and Ganger, Gregory R. and Gibbons, Phillip B. and Zaharia, Matei},
title = {PipeDream: generalized pipeline parallelism for DNN training},
year = {2019},
isbn = {9781450368735},
publisher = {Association for Computing Machinery},
address = {New York, NY, USA},
url = {https://doi.org/10.1145/3341301.3359646},
doi = {10.1145/3341301.3359646},
booktitle = {Proceedings of the 27th ACM Symposium on Operating Systems Principles},
pages = {1–15},
numpages = {15},
location = {Huntsville, Ontario, Canada},
series = {SOSP '19}
}

@misc{bonato2026spritzpathawareloadbalancing,
      title={Spritz: Path-Aware Load Balancing in Low-Diameter Networks}, 
      author={Tommaso Bonato and Ales Kubicek and Abdul Kabbani and Ahmad Ghalayini and Maciej Besta and Torsten Hoefler},
      year={2026},
      eprint={2602.19567},
      archivePrefix={arXiv},
      primaryClass={cs.NI},
      url={https://arxiv.org/abs/2602.19567}, 
}

@inproceedings {themis,
author = {Kshiteej Mahajan and Arjun Balasubramanian and Arjun Singhvi and Shivaram Venkataraman and Aditya Akella and Amar Phanishayee and Shuchi Chawla},
title = {Themis: Fair and Efficient {GPU} Cluster Scheduling },
booktitle = {17th USENIX Symposium on Networked Systems Design and Implementation (NSDI 20)},
year = {2020},
isbn = {978-1-939133-13-7},
address = {Santa Clara, CA},
pages = {289--304},
url = {https://www.usenix.org/conference/nsdi20/presentation/mahajan},
publisher = {USENIX Association},
month = feb
}

@inproceedings{siriusGPUsharing,
author = {Wang, Jiali and Wang, Yankui and Han, Mingcong and Chen, Rong},
title = {Colocating ML inference and training with fast GPU memory handover},
year = {2025},
isbn = {978-1-939133-48-9},
publisher = {USENIX Association},
address = {USA},
booktitle = {Proceedings of the 2025 USENIX Conference on Usenix Annual Technical Conference},
articleno = {98},
numpages = {19},
location = {Boston, MA, USA},
series = {USENIX ATC '25}
}

@inproceedings{TacosWon_2024,
   title={TACOS: Topology-Aware Collective Algorithm Synthesizer for Distributed Machine Learning},
   url={http://dx.doi.org/10.1109/MICRO61859.2024.00068},
   DOI={10.1109/micro61859.2024.00068},
   booktitle={2024 57th IEEE/ACM International Symposium on Microarchitecture (MICRO)},
   publisher={IEEE},
   author={Won, William and Elavazhagan, Midhilesh and Srinivasan, Sudarshan and Gupta, Swati and Krishna, Tushar},
   year={2024},
   month=Nov, pages={856–870} }

@article{ShuttleDispatching2021,
author = {Tafreshian, Amir and Abdolmaleki, Mojtaba and Masoud, Neda and Wang, Huizhu},
year = {2021},
month = {06},
pages = {227-259},
title = {Proactive Shuttle Dispatching in Large-Scale Dynamic Dial-a-Ride Systems},
volume = {150},
journal = {Transportation Research Part B Methodological},
doi = {10.1016/j.trb.2021.06.002}
}

@article{BELIERES2021102203,
title = {A time-expanded network reduction matheuristic for the logistics service network design problem},
journal = {Transportation Research Part E: Logistics and Transportation Review},
volume = {147},
pages = {102203},
year = {2021},
issn = {1366-5545},
doi = {https://doi.org/10.1016/j.tre.2020.102203},
url = {https://www.sciencedirect.com/science/article/pii/S1366554520308450},
author = {Simon Belieres and Mike Hewitt and Nicolas Jozefowiez and Frédéric Semet}
}

@inproceedings{tenFlowDep2002,
author = {K\"{o}hler, Ekkehard and Langkau, Katharina and Skutella, Martin},
title = {Time-Expanded Graphs for Flow-Dependent Transit Times},
year = {2002},
isbn = {3540441808},
publisher = {Springer-Verlag},
address = {Berlin, Heidelberg},
booktitle = {Proceedings of the 10th Annual European Symposium on Algorithms},
pages = {599–611},
numpages = {13},
series = {ESA '02}
}

@inproceedings{UCMFastRDMA,
author = {Shen, Huijun and Yang, Jian and Yue, Zelong and Guo, Xingyu and Yin, Xijin and An, Lang and Chen, Yulin and Ding, Jie and Wu, Hongyu and Zhang, Yong and Ye, Jianxi and Chen, Guo},
title = {UCM: Fast and Maintainable User-space RDMA Connection Setup},
year = {2025},
isbn = {9798400714016},
publisher = {Association for Computing Machinery},
address = {New York, NY, USA},
url = {https://doi.org/10.1145/3735358.3735388},
doi = {10.1145/3735358.3735388},
booktitle = {Proceedings of the 9th Asia-Pacific Workshop on Networking},
pages = {24–30},
numpages = {7},
location = {
},
series = {APNET '25}
}

@misc{ncclx-meta,
      title={Collective Communication for 100k+ GPUs}, 
      author={Min Si and Pavan Balaji and Yongzhou Chen and Ching-Hsiang Chu and Adi Gangidi and Saif Hasan and Subodh Iyengar and Dan Johnson and Bingzhe Liu and Regina Ren and Deep Shah and Ashmitha Jeevaraj Shetty and Greg Steinbrecher and Yulun Wang and Bruce Wu and Xinfeng Xie and Jingyi Yang and Mingran Yang and Kenny Yu and Minlan Yu and Cen Zhao and Wes Bland and Denis Boyda and Suman Gumudavelli and Prashanth Kannan and Cristian Lumezanu and Rui Miao and Zhe Qu and Venkat Ramesh and Maxim Samoylov and Jan Seidel and Srikanth Sundaresan and Feng Tian and Qiye Tan and Shuqiang Zhang and Yimeng Zhao and Shengbao Zheng and Art Zhu and Hongyi Zeng},
      year={2026},
      eprint={2510.20171},
      archivePrefix={arXiv},
      primaryClass={cs.DC},
      url={https://arxiv.org/abs/2510.20171}, 
}

@inproceedings{alibaba-hpn,
author = {Qian, Kun and Xi, Yongqing and Cao, Jiamin and Gao, Jiaqi and Xu, Yichi and Guan, Yu and Fu, Binzhang and Shi, Xuemei and Zhu, Fangbo and Miao, Rui and Wang, Chao and Wang, Peng and Zhang, Pengcheng and Zeng, Xianlong and Ruan, Eddie and Yao, Zhiping and Zhai, Ennan and Cai, Dennis},
title = {Alibaba HPN: A Data Center Network for Large Language Model Training},
year = {2024},
isbn = {9798400706141},
publisher = {Association for Computing Machinery},
address = {New York, NY, USA},
url = {https://doi.org/10.1145/3651890.3672265},
doi = {10.1145/3651890.3672265},
booktitle = {Proceedings of the ACM SIGCOMM 2024 Conference},
pages = {691–706},
numpages = {16},
location = {Sydney, NSW, Australia},
series = {ACM SIGCOMM '24}
}

@inproceedings{plbSigcomm2022,title	= {PLB: Congestion Signals are Simple and Effective for Network Load Balancing},author	= {Abdul Kabbani and David J. Wetherall and Gautam Kumar and Junhua Yan and Kira Yin and Masoud Moshref and Mubashir Adnan Qureshi and Qiaobin Fu and Van Jacobson and Yuchung Cheng},year	= {2022}}

@misc{mcclure2026loadbalancingaitraining,
      title={Load Balancing for AI Training Workloads}, 
      author={Sarah McClure and Evyatar Cohen and Alex Shpiner and Mark Silberstein and Sylvia Ratnasamy and Scott Shenker and Isaac Keslassy},
      year={2026},
      eprint={2507.21372},
      archivePrefix={arXiv},
      primaryClass={cs.NI},
      url={https://arxiv.org/abs/2507.21372}, 
}

@inproceedings{congPatternsMetaIMC2025,
author = {Ghorbani, Soudeh and Zhao, Yimeng and Sundaresan, Srikanth and Zhang, Ying and Zeng, Yijing and Sharma, Abhigyan and Kannan, Prashanth and Lumezanu, Cristian},
title = {Congestion Patterns in a Large-scale RDMA Datacenter},
year = {2025},
isbn = {9798400718601},
publisher = {Association for Computing Machinery},
address = {New York, NY, USA},
url = {https://doi.org/10.1145/3730567.3764494},
doi = {10.1145/3730567.3764494},
booktitle = {Proceedings of the 2025 ACM Internet Measurement Conference},
pages = {944–951},
numpages = {8},
location = {USA},
series = {IMC '25}
}

@inproceedings{meta-rdma,
author = {Gangidi, Adithya and Miao, Rui and Zheng, Shengbao and Bondu, Sai Jayesh and Goes, Guilherme and Morsy, Hany and Puri, Rohit and Riftadi, Mohammad and Shetty, Ashmitha Jeevaraj and Yang, Jingyi and Zhang, Shuqiang and Fernandez, Mikel Jimenez and Gandham, Shashidhar and Zeng, Hongyi},
title = {RDMA over Ethernet for Distributed Training at Meta Scale},
year = {2024},
isbn = {9798400706141},
publisher = {Association for Computing Machinery},
address = {New York, NY, USA},
url = {https://doi.org/10.1145/3651890.3672233},
doi = {10.1145/3651890.3672233},
booktitle = {Proceedings of the ACM SIGCOMM 2024 Conference},
pages = {57–70},
numpages = {14},
location = {Sydney, NSW, Australia},
series = {ACM SIGCOMM '24}
}

@inproceedings {msft-usenix,
author = {Myeongjae Jeon and Shivaram Venkataraman and Amar Phanishayee and Junjie Qian and Wencong Xiao and Fan Yang},
title = {Analysis of {Large-Scale} {Multi-Tenant} {GPU} Clusters for {DNN} Training Workloads},
booktitle = {2019 USENIX Annual Technical Conference (USENIX ATC 19)},
year = {2019},
isbn = {978-1-939133-03-8},
address = {Renton, WA},
pages = {947--960},
url = {https://www.usenix.org/conference/atc19/presentation/jeon},
publisher = {USENIX Association},
month = jul
}

@Inbook{ns3,
author="Riley, George F.
and Henderson, Thomas R.",
editor="Wehrle, Klaus
and G{\"u}ne{\c{s}}, Mesut
and Gross, James",
title="The ns-3 Network Simulator",
bookTitle="Modeling and Tools for Network Simulation",
year="2010",
publisher="Springer Berlin Heidelberg",
address="Berlin, Heidelberg",
pages="15--34",
isbn="978-3-642-12331-3",
doi="10.1007/978-3-642-12331-3_2",
url="https://doi.org/10.1007/978-3-642-12331-3_2"
}

@misc{nccl-collectives-2,
  author       = {{NVIDIA}},
  title        = {{NCCL Documentation: Collective Operations}},
  howpublished = {\url{https://docs.nvidia.com/deeplearning/nccl/user-guide/docs/usage/collectives.html}},
  year         = {2026},
  note         = {Accessed: 2026-04-23}
}

@inproceedings{msft-gao,
author = {Gao, Yanjie and He, Yichen and Li, Xinze and Zhao, Bo and Lin, Haoxiang and Liang, Yoyo and Zhong, Jing and Zhang, Hongyu and Wang, Jingzhou and Zeng, Yonghua and Gui, Keli and Tong, Jie and Yang, Mao},
title = {An Empirical Study on Low GPU Utilization of Deep Learning Jobs},
year = {2024},
isbn = {9798400702174},
publisher = {Association for Computing Machinery},
address = {New York, NY, USA},
url = {https://doi.org/10.1145/3597503.3639232},
doi = {10.1145/3597503.3639232},
booktitle = {Proceedings of the IEEE/ACM 46th International Conference on Software Engineering},
articleno = {96},
numpages = {13},
location = {Lisbon, Portugal},
series = {ICSE '24}
}
